\documentclass[aps,twocolumn,superscriptaddress,a4paper]{revtex4}

\usepackage{graphicx}
\usepackage{amsmath}
\usepackage{amssymb}
\usepackage{bm}

\newcommand{\unites}[1]{\; \mathrm{#1}}   

\begin{document}
\title{Nonadiabatic effects in the double ionization of atoms driven by a circularly polarized laser pulse}
%\date{\today}

\author{J. Dubois}
\affiliation{Aix Marseille Univ, CNRS, Centrale Marseille, I2M, Marseille, France}
\affiliation{Max Planck Institute for the Physics of Complex Systems, Dresden, Germany}
\author{C. Chandre}
\affiliation{Aix Marseille Univ, CNRS, Centrale Marseille, I2M, Marseille, France}
\author{T. Uzer}
\affiliation{School of Physics, Georgia Institute of Technology, Atlanta, Georgia 30332-0430, USA}

\begin{abstract}
We study the double ionization of atoms subjected to circularly polarized (CP) laser pulses. We analyze two fundamental ionization processes: the sequential (SDI) and non-sequential (NSDI) double ionization in the light of the rotating frame (RF) which naturally embeds nonadiabatic effects in CP pulses. We use and compare two adiabatic approximations: The adiabatic approximation in the laboratory frame (LF) and the adiabatic approximation in the RF. The adiabatic approximation in the RF encapsulates the energy variations of the electrons on subcycle timescales happening in the LF and this, by fully taking into account the ion-electron interaction. This allows us to identify two nonadiabatic effects including the lowering of the threshold intensity at which over-the-barrier ionization happens and the lowering of the ionization time of the electrons. As a consequence, these nonadiabatic effects facilitate over-the-barrier ionization and recollision-induced ionizations. We analyze the outcomes of these nonadiabatic effects on the recollision mechanism. We show that the laser envelope plays an instrumental role in a recollision channel in CP pulses at the heart of NSDI.
\end{abstract}

% PhySH
% Atomic and Molecular Collisions
% Hamiltonian Systems
% Chaos and Nonlinear Dynamics

\maketitle

\section{Introduction}
Intense laser pulses generate forces which are potentially comparable to Coulomb forces inside atoms and molecules. Driven by such intense pulses, atoms and molecules typically ionize or multiple ionize, depending on the parameters of the laser pulse (typically its intensity and its wavelength). Circularly polarized (CP) laser pulses have been instrumental since they offer a drastically different picture of the laser-matter interaction than afforded by linear polarization. As a prominent example, attoclock~\citep{Hofmann2019} aims at timing the release of the ionized electrons and probing the ionization processes at the natural timescale of the electron~\citep{Eckle2008, Pfeiffer2011, Torlina2015, Ni2016, Eckart2018_PRL, Eckart2018_NatPhys, Sainadh2019}--the attosecond. In attoclock, CP or near-CP laser pulses are employed to isolate, for instance, nonadiabatic effects induced by the subcycle temporal variations of the laser pulse~\citep{Klaiber2015, Ni2018, Eckart2018_PRL}. The assumption behind this setup is the absence of recollisions in CP pulses, which is supported by the strong sideways drift momentum of the electron after ionization pushing it away from the parent ion after ionization~\citep{Corkum1993}. The absence of recollisions allows for clear photoelectron momentum distributions (PMDs) in order to probe Coulomb effects~\citep{Sainadh2019}. 
\par 
For double ionization (DI), the attoclock setup is a natural way for probing the electron-electron correlations in atoms~\cite{Bryan2006, Pfeiffer2011}. The double ionization processes occur in mainly two categories: a sequential double ionization (SDI) channel and a nonsequential double ionization (NSDI) one.  
In SDI, the electrons are ripped out of the core one after the other independently. Understanding the conditions under which the electrons ionize independently from each other allows the assessment of the contribution from electron-electron correlations in experiments~\cite{Pfeiffer2011}. 
In contrast to SDI, NSDI is when two electrons are ripped out of the core with a significant contribution of electron-electron correlations. As a consequence of NSDI, the DI probability can be enhanced by several orders of magnitude compared to the SDI probability~\cite{Corkum1993, Walker1994}. When the DI probability is depicted as a function of the laser intensity, a knee structure may be observed, as a signature of NSDI~\citep{Walker1994,Ho2005, Mauger2009} and as a hallmark of the prominent role of the electron-electron correlation. These manifestations of electron-electron correlations have been routinely observed for decades now, but mainly for linearly or close-to-linearly polarized fields. 
\par 
In recent years, there is growing evidence that recollision-driven NSDI can be observed in CP fields for selected parameters of the laser field~\cite{Mauger2010_PRL, Wang2010_PRL, Fu2012, Hang2020}. The role and importance of electron-electron correlations triggered by a CP laser field have to be clearly assessed as parameters of the laser are varied, in order to properly identify the correct channels for double ionization.  
\par 
In this article, we study SDI and NSDI processes in atoms subjected to CP pulses. The conditions under which the two electrons ionize independently and the conditions under which a pre-ionized electron returns to the parent ion are investigated in the framework of the single-active electron (SAE) approximation~\cite{Pfeiffer2011, Wang2012}.
In the SAE approximation, a first difficulty arises due to the resonances of the almost time-periodic laser field with the electron dynamics which increase its energy on subcycle timescales, giving rise to what is referred to as nonadiabatic effects~\cite{Barth2011,Barth2014}. We move the combined Coulomb-time-dependent field system to a rotating frame~\cite{Mauger2010_PRL, Kamor2013} (RF), where the energy of the electron does not vary on these subcycle timescales. We use and compare two adiabatic approximations: The adiabatic approximation in the laboratory frame (LF) and the adiabatic approximation in the RF. The adiabatic approximation in the RF naturally embeds the energy variations of the electron on subcycle timescales happening in the LF, and thus, by fully taking into account the ion-electron interactions. Here this allows us to identify two nonadiabatic effects in the ionization process (compared with the adiabatic approximation): The lowering of the threshold intensity for which over-the-barrier ionization occurs, and the lowering of the ionization time of the electrons in the over-the-barrier ionization regime. One of our conclusions is that the NSDI observed in experiments~\cite{Hang2020,Gillen2001} occurs for intensities in the regime of over-the-barrier ionization. Our result departs from the claim that NSDI happens in the tunneling regime when considering the adiabatic approximation~\cite{Fu2012}. One of the major advantages of over-the-barrier ionization is that it lends itself very well to an accurate description by classical mechanics or more precisely by nonlinear dynamics. We take advantage of the classical trajectories of the electron and their analysis to derive the conditions under which an electron can recollide in CP pulses. In addition, the RF allows us to focus our analysis on a larger timescale, on the order of the timescale of the laser envelope, allowing us to provide the specific conditions for a recollision channel identified in Ref.~\cite{Dubois2020}. We show the instrumental role played by the laser envelope on these conditions.
\par
In Sec.~\ref{sec:Hamitlonian_model}, we introduce the Hamiltonian models and DI probability curves as a function of the laser intensity. We derive two models in the single-active electron approximation, which we use for analyzing the double ionization processes throughout the article. These models are provided in the LF and in the RF. In Sec.~\ref{sec:analysis}, we provide a brief analysis on their effective potentials, and we identify the prominent role of two fixed points: a stable point at the bottom of the potential well and a saddle point at the top of the barrier induced by the laser field. In Sec.~\ref{sec:SDI_mechanisms}, we investigate the role of the saddle point on the ionization times in the SDI regime. We derive the conditions under which over-the-barrier ionization occurs and the release times of the electrons as a function of the laser and atomic parameters by accounting for the electron energy variations on subcycle timescales happening in the RF. We compare our results with experiments where no recollisions are observed. In Sec.~\ref{sec:NSDI_mechanisms}, with the help of the results derived in Sec.~\ref{sec:SDI_mechanisms}, we derive the conditions under which recollisions can be observed in CP pulses. 

\section{Hamiltonian models for the double ionization \label{sec:Hamitlonian_model}}

\begin{figure}
	\centering
	\includegraphics[width=0.5\textwidth]{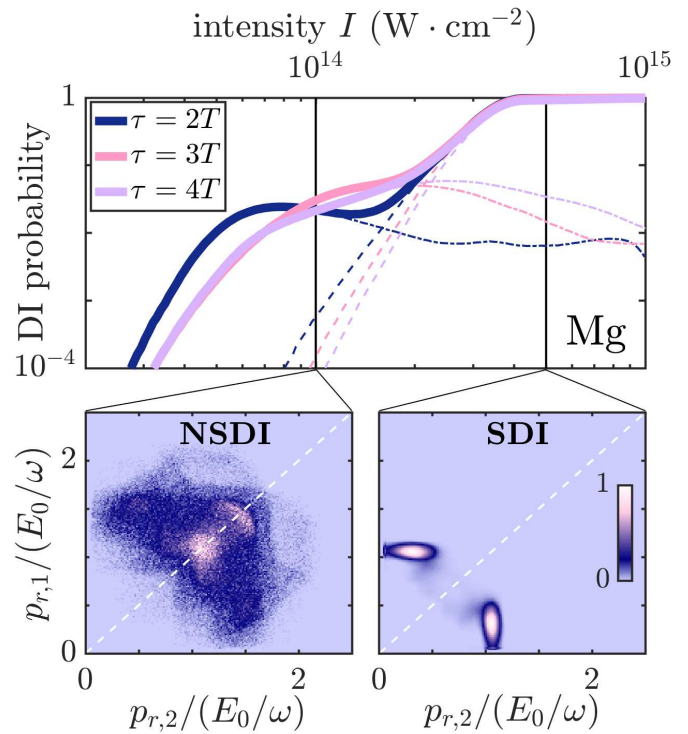}
	\caption{Double ionization (DI) probability of Hamiltonian~\eqref{eq:Hamiltonian_NSDI_LF} in 2D as a function of the laser intensity $I$ (solid thick curves) for $\mathrm{Mg}$ ($\mathcal{E}_g = - 0.83$, $a=3$) and CP pulses. The laser envelope is trapezoidal with duration of ramp-up $\tau$, plateau $6T - \tau$, and ramp-down $2T$. The blue, pink and purple curves are for $\tau = 2T$, $3T$ and $4T$, respectively. The thin dashed-dotted and dashed curves are the probability of NSDI and SDI, respectively. The lower panels are the correlated photoelectron radial momentum distributions for $\tau = 2 T$ with $p_{r,k} = \mathbf{r}_k \cdot \mathbf{p}_k /|\mathbf{r}_k|$ for $k=1,2$. Left and right panel are for $I =  10^{14} \; \mathrm{W}\cdot \mathrm{cm}^{-2}$ and $I = 5 \times 10^{14} \; \mathrm{W}\cdot \mathrm{cm}^{-2}$ (indicated by the black vertical lines on the top panel), respectively. The dashed white lines is when $p_{r,1} = p_{r,2}$.}
	\label{fig:probability_curves}
\end{figure}

\begin{figure}
	\centering
	\includegraphics[width=0.5\textwidth]{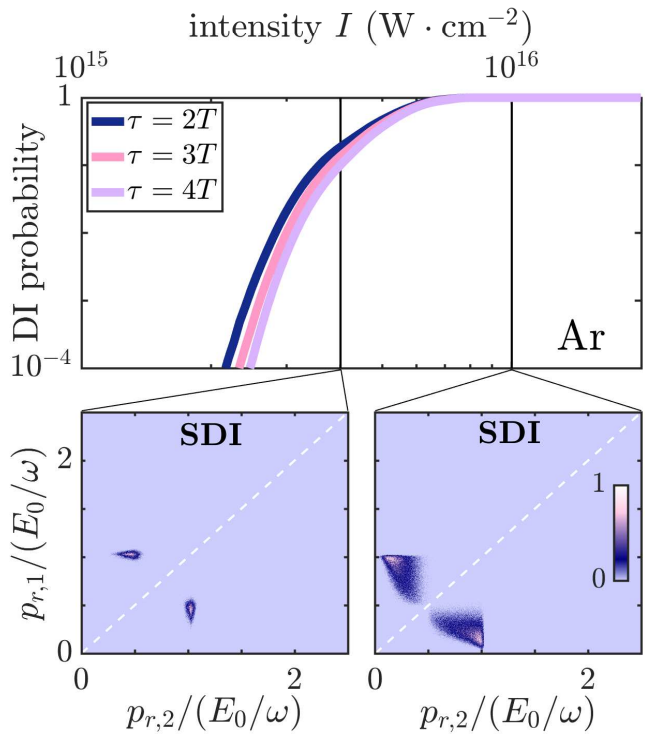}
	\caption{Double ionization (DI) probability of Hamiltonian~\eqref{eq:Hamiltonian_NSDI_LF} in 2D as a function of the laser intensity $I$ (solid thick curves) for $\mathrm{Ar}$ ($\mathcal{E}_g = - 1.60$, $a=1.5$) and CP pulses. The laser envelope is trapezoidal with duration of ramp-up $\tau$, plateau $6T - \tau$, and ramp-down $2T$. The blue, pink and purple curves are for $\tau = 2T$, $3T$ and $4T$, respectively. The lower panels are the correlated photoelectron radial momentum distributions for $\tau = 2 T$ with $p_{r,k} = \mathbf{r}_k \cdot \mathbf{p}_k /|\mathbf{r}_k|$ for $k=1,2$. Left and right panels are for $I =  4\times 10^{15} \; \mathrm{W}\cdot \mathrm{cm}^{-2}$ and $I =  10^{16} \; \mathrm{W}\cdot \mathrm{cm}^{-2}$ (indicated by the black vertical lines on the top panel), respectively. The dashed lines is when $p_{r,1} = p_{r,2}$.}
	\label{fig:probability_curves_Ar}
\end{figure}

We use a fully classical model in order to describe the double ionization processes in atoms subjected to CP laser pulses. Atomic units ($\mathrm{a.u.}$) are used unless stated otherwise. The laser wavelength is $780\; \mathrm{nm}$ (corresponding to a laser frequency $\omega = 0.0584$), unless stated otherwise. 

\subsection{Two-active electron model \label{sec:two_active_electrons_probability_curves}}

First, we consider a two-active electron system driven by a CP laser pulse. In the LF of basis vectors $(\hat{\mathbf{x}} , \hat{\mathbf{y}} , \hat{\mathbf{z}} )$, the laser pulse reads 
$$
\mathbf{E} (t) = E_0 f(t) [ \hat{\mathbf{x}} \cos ( \omega t ) + \hat{\mathbf{y}}  \sin ( \omega t ) ] . 
$$
The laser amplitude $E_0$ is related to its intensity $I$ by the relation $E_0 = \sqrt{I/2}$. We recall that in SI units, the intensity is $I  \; [\mathrm{W}\cdot \mathrm{cm}^{-2}] = 3.5 \times 10^{16} I \; [\mathrm{a.u.}]$. In this manuscript, we consider a trapezoidal envelope with a ramp-up of duration $\tau$, a plateau of duration $6 T - \tau$, and a ramp-down of duration $2T$ where $T=2\pi/\omega$, unless stated otherwise. The two-active electrons are labeled by $k=1$ and $2$, with positions and momenta $\mathbf{r}_k$ and $\mathbf{p}_k$, respectively. We denote $\mathrm{e}_k^-$ as the $k$-th electron. We consider the following Hamiltonian for the two-active electron system in the dipole approximation~\citep{Mauger2010_PRL}:
\begin{eqnarray}
\label{eq:Hamiltonian_NSDI_LF}
&& H \left( \mathbf{r}_1 , \mathbf{p}_1 ,\mathbf{r}_2 , \mathbf{p}_2  , t \right) = \nonumber \\ 
&& \sum_{k=1}^2 \left[ \dfrac{| \mathbf{p}_k|^2}{2} - \dfrac{2}{\sqrt{|\mathbf{r}_k|^2 + a^2}}  + \mathbf{r}_k  \cdot \mathbf{E}(t) \right] \nonumber \\ 
	&&+ \dfrac{1}{\sqrt{| \mathbf{r}_1 - \mathbf{r}_2|^2+1}} .
\end{eqnarray}
We use soft Coulomb potentials~\citep{Javanainen1988} to describe the ion-electron and the electron-electron interactions (see also Refs.~\citep{Ho2005,Mauger2009,Mauger2010_PRL}). In our simulations, we consider an ensemble of initial conditions for Hamiltonian~\eqref{eq:Hamiltonian_NSDI_LF}, generated randomly using a microcanonical distribution such that $H (\mathbf{r}_1 , \mathbf{p}_1 ,\mathbf{r}_2 , \mathbf{p}_2 , 0) = \mathcal{E}_g$, where $\mathcal{E}_g$ is the ground-state energy of the two-active electron system (obtained as the sum of the first two ionization potentials of the corresponding atom).
In what follows, we mainly focus on two atoms: Mg where the ground-state energy is $\mathcal{E}_g = - 0.83$, and Ar where the ground-state energy is $\mathcal{E}_g = - 1.60$. In order to avoid self-ionization and non-emptiness of the ground state, we choose the following softening parameters: $a = 3$ for Mg and $a = 1.5$ for Ar. We integrate numerically the trajectories to determine which ones lead to double ionization or not. 
In Fig.~\ref{fig:probability_curves}, the thick curves are the DI probability curves of Hamiltonian~\eqref{eq:Hamiltonian_NSDI_LF} in 2D for Mg for different ramp-up durations $\tau$, as functions of the intensity. In Fig.~\ref{fig:probability_curves_Ar}, the same curves are done for Ar using the same laser envelopes and laser frequency as in Fig.~\ref{fig:probability_curves} for the laser field. 
\par
For all the three ramp-up durations, the DI probability curves exhibit a knee structure for Mg, a hallmark of electron-electron correlation in the double ionization process. A bump is observed for intensities  around $3\times 10^{13}$ to $2\times 10^{14}$ $ \mathrm{W}\cdot \mathrm{cm}^{-2}$. Then, for intensities between $2\times 10^{14}$ and $4.5\times 10^{14}$ $ \mathrm{W}\cdot \mathrm{cm}^{-2}$, the DI probability increases monotonically with intensity, and saturates around $100\%$ for intensities $I \gtrsim 4.5\times 10^{14} \; \mathrm{W}\cdot \mathrm{cm}^{-2}$.  
\par
In order to analyze these DI probability curves, we distinguish the SDI and the NSDI among the DIs. We consider a trajectory to be an NSDI when there exists $t_{\rm i}$ and $t_{\rm r}$ such that: one electron leaves the core at time $t_{\rm i}$, e.g., $|\mathbf{r}_1 (t_{\rm i})| > R$, then comes back to the core at time $t_{\rm r}$ while the other electron remains bounded, i.e., $|\mathbf{r}_1 (t_{\rm r})| < R$ and $|\mathbf{r}_2 (t)| < R$ for $t \in [t_{\rm i},t_{\rm r}]$, and finally both ionize, i.e., for sufficiently large $t$, $|\mathbf{r}_{k} (t)| > 100$ for $k=1$ and $2$. Throughout the manuscript, the threshold for pre-ionization and return is $R = 5$. In Fig.~\ref{fig:probability_curves}, the thin dashed and thin dash-dotted curves are the SDI and NSDI probability curves, respectively. We observe that the bump in the DI probability curves is dominated by the NSDI, while the high-intensity part is dominated by the SDI. The knee structure is a manifestation of a crossover between NSDI and SDI as intensity increases. Also, noteworthy is the presence of NSDI for large values of the intensities (even when the double ionization is saturated), even though the dominant process is overwhelmingly SDI. 
\par
For $I \gtrsim 10^{14} \; \mathrm{W} \cdot \mathrm{cm}^{-2}$, the SDI probability curves (thin dashed curves) monotonically increases for increasing intensity. This behavior is also observed in Fig.~\ref{fig:probability_curves_Ar} for Ar for which there is only SDI. It is important to notice that the SDI probability depends weakly on the laser envelope: For instance, we observe that the slope of the SDI probability curve does not change significantly with the duration of the ramp-up, and these probabilities all saturate around the same intensity $I \sim 4.5\times 10^{14} \; \mathrm{W}\cdot \mathrm{cm}^{-2}$ for Mg, and around $I \sim 8\times 10^{15} \; \mathrm{W}\cdot \mathrm{cm}^{-2}$ for Ar. In contrast, we observe that the NSDI probability curves (thin dash-dotted curves) are strongly influenced by the laser envelope: For instance, the location and the shape of the main bumps vary significantly with $\tau$, especially between $\tau = 2T$ and $\tau = 3T$. However, it seems that there is a rather small dependence between $\tau = 3T$ and $\tau = 4T$.
\par
One of our objectives is to understand the main features of SDI and NSDI mechanisms in CP laser pulses, and in particular their strong dependence with the laser envelope. In order to proceed, we consider reduced models whose dynamics can be more easily analyzed.  

\subsection{Single-active electron models}

In order to analyze the dynamics associated with Hamiltonian~\eqref{eq:Hamiltonian_NSDI_LF}, we use single-active electron (SAE) approximations~\citep{Wang2012,Lan2014,Kamor2013}, also referred to as inner/outer electron approximations. We consider that the electrons ionize one after the other. Without loss of generality, we consider $\mathrm{e}_1^-$ as the outer electron, which ionizes first, and $\mathrm{e}_2^-$ as the inner electron, which ionizes second. When the outer electron is outside the core region, we have $| \mathbf{r}_1 -  \mathbf{r}_2 | \approx | \mathbf{r}_1| \gg a^2$. It translates into
$$
- \dfrac{2}{\sqrt{| \mathbf{r}_1 |^2 + a^2}} + \dfrac{1}{\sqrt{|  \mathbf{r}_1 -  \mathbf{r}_2|^2 + 1}} \approx - \dfrac{1}{\sqrt{| \mathbf{r}_1|^2 + a^2}} .
$$
As a consequence, the nuclear charge seen by the outer electron is screened by the inner electron. When the two electrons are far from each other, Hamiltonian~\eqref{eq:Hamiltonian_NSDI_LF} can be written in terms of a sum of two uncoupled Hamiltonians
\begin{equation}
\label{eq:Hamiltonian_NSDI_screen_effect}
H ( \mathbf{r}_1, \mathbf{p}_1 , \mathbf{r}_2 , \mathbf{p}_2 , t ) \approx H_1 ( \mathbf{r}_1, \mathbf{p}_1,t) + H_2 ( \mathbf{r}_2, \mathbf{p}_2,t) ,
\end{equation}
where Hamiltonians $H_1$ and $H_2$ describe the outer and inner electron dynamics, respectively. The reduced Hamiltonians are
\begin{equation}
\label{eq:Hamiltonian_LF_1ae}
H_k ( \mathbf{r}, \mathbf{p}, t) = \dfrac{| \mathbf{p}|^2}{2} + V_k ( \mathbf{r}) + \mathbf{r} \cdot \mathbf{E} (t),
\end{equation}
where $V_k ( \mathbf{r}) = - Z_k / \sqrt{|\mathbf{r}|^2 + a^2}$.
The effective charge of the outer electron is $Z_1 = 1$, and the effective charge of the inner electron is $Z_2 = 2$.

\subsection{Rotating frame}
For CP laser pulses, it is more convenient to analyze the dynamics in a frame which is rotating with the laser field. It is worth noticing that the dynamics in the LF and in the RF are linked by an invertible change of coordinates, so there is the same amount of information obtained from one framework or another. We argue that the dynamics in the RF displays this information more clearly. The full advantages of seeing the dynamics in the RF are clearly visible when making another approximation, namely the adiabatic approximation. For instance, it highlights the specific role played by the laser envelope in ionization and recollision processes. This will be detailed in the next section. 
\par 
We rewrite the electron dynamics in a rotating frame (RF) of basis vectors $(\hat{\tilde{\mathbf{x}}}, \hat{\tilde{\mathbf{y}}}, \hat{{\mathbf{z}}})$. In this RF, the CP laser pulse is unidirectional and does no longer depends on the fast oscillations of the field with frequency $\omega$. The canonical change of coordinates from the LF to the RF is $\mathbf{r} =  {R}(\omega t) \tilde{\mathbf{r}}$ and $\mathbf{p} =  {R}(\omega t) \tilde{\mathbf{p}}$. The matrix ${R}$ is a rotation matrix in the polarization plane (i.e., of axis $\hat{\mathbf{z}}$) with angle $\omega t$. In the RF, Hamiltonian~\eqref{eq:Hamiltonian_NSDI_LF} becomes
\begin{eqnarray}
\label{eq:Hamiltonian_NSDI_RF}
&& \tilde{H} \left( \tilde{\mathbf{r}}_1 , \tilde{\mathbf{p}}_1 ,\tilde{\mathbf{r}}_2 , \tilde{\mathbf{p}}_2  , t \right) = \nonumber \\ 
&& \sum_{k=1}^2 \left[ \dfrac{| \tilde{\mathbf{p}}_k|^2}{2}- \omega \tilde{\mathbf{r}}_k \times \tilde{\mathbf{p}}_k \cdot \hat{{\mathbf{z}}} - \dfrac{2}{\sqrt{|\tilde{\mathbf{r}}_k|^2 + a^2}}  + \tilde{\mathbf{r}}_k  \cdot \tilde{\mathbf{E}}(t) \right] \nonumber \\ 
	&&+ \dfrac{1}{\sqrt{| \tilde{\mathbf{r}}_1 - \tilde{\mathbf{r}}_2|^2+1}}, 
\end{eqnarray}
with $\tilde{\mathbf{E}} (t) = f(t) E_0 \hat{\tilde{\mathbf{x}}}$. The term $-\omega \tilde{\mathbf{r}} \times \tilde{\mathbf{p}} \cdot \hat{{\mathbf{z}}}$ corresponds to the Coriolis potential (as a result of the time-dependence of the canonical change of coordinates). The SAE
Hamiltonians~\eqref{eq:Hamiltonian_LF_1ae} in the RF become
\begin{equation}
\label{eq:Hamiltonian_RF_1ae}
\tilde{H}_k ( \tilde{\mathbf{r}} , \tilde{\mathbf{p}}, t) = \dfrac{| \tilde{\mathbf{p}}|^2}{2} - \omega \tilde{\mathbf{r}} \times \tilde{\mathbf{p}} \cdot \hat{{\mathbf{z}}} + V_k ( \tilde{\mathbf{r}} ) + \tilde{\mathbf{r}} \cdot \tilde{\mathbf{E}}(t) .
\end{equation}
We notice that for rotationally invariant ionic core potentials, which is the case for atomic potentials, and in particular for the soft-Coulomb potential, the Hamiltonians in the RF do no longer depend on the carrier-envelope phase of the laser. Consequently, quantities such as for instance DI probabilities do not depend on the carrier-envelope phase for rotationally invariant initial distributions of the initial conditions~\citep{Xu2015}. 

\section{Analysis of the effective SAE potentials}
\label{sec:analysis}

\begin{figure*}
	\centering
	\includegraphics[width=.9\textwidth]{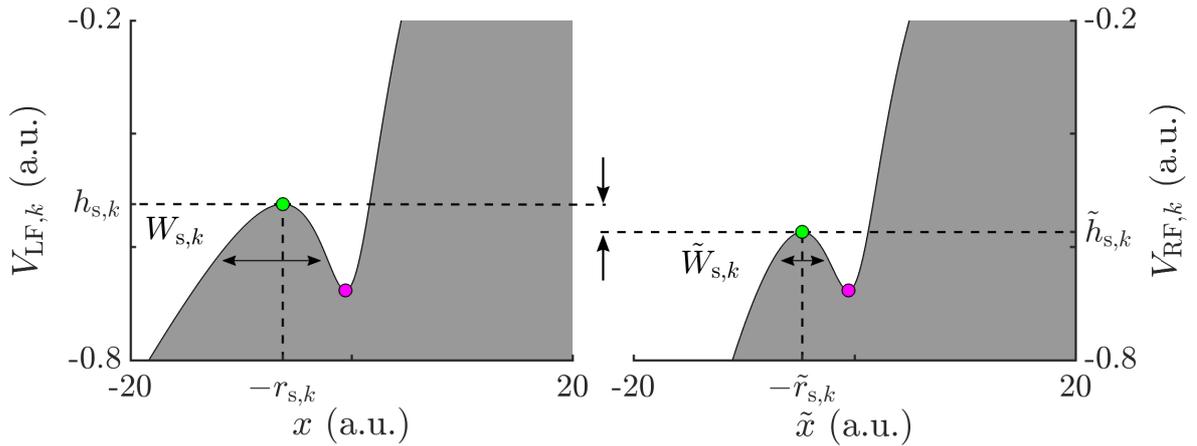}
	\caption{Effective potentials in the LF given by Eq.~\eqref{eq:Veff_LF} (left panel) and the RF given by Eq.~\eqref{eq:Veff_RF} (right panel) for $I = 10^{14} \; \mathrm{W}\cdot\mathrm{cm}^{-2}$, $f=1$, $a=3$, $k=2$ and $\omega t = 0$ (laser field along the $\hat{\mathbf{x}}$-axis in the LF). The green dots indicate the location of the saddle points in the LF $(- r_{{\rm s},k} , h_{{\rm s},k})$ and in the RF $(- \tilde{r}_{{\rm s},k} , \tilde{h}_{{\rm s},k})$. The magenta dots indicate the location of the stable points in the LF $(- r_{{\rm b},k} , h_{{\rm b},k})$ and in the RF $(- \tilde{r}_{{\rm b},k} , \tilde{h}_{{\rm b},k})$. The double arrows indicate the width of the barrier of the effective potentials in the LF, $W_{{\rm s},k}$ and in the RF, $\tilde{W}_{{\rm s},k}$.}
	\label{fig:Veff_LF_RF}
\end{figure*}

In both the LF and the RF, the Hamiltonians in the SAE approximation are given by a kinetic term plus an effective potential. These effective potentials are all characterized by the existence of a stable point at the bottom of the potential well, ensuring that an electron close to the ionic core remains bounded, and by a saddle point at the top of the barrier induced by the laser field, allowing for ionization when the electron has enough energy. Below, we analyze these two critical points in the adiabatic approximation in the LF and in the adiabatic approximation in the RF, according to the parameters of the laser field.  

\subsection{Adiabatic approximation in the LF}
In Hamiltonians~\eqref{eq:Hamiltonian_LF_1ae}, we observe that the time-dependent term is almost $T$-periodic. As a consequence, the electron energy varies on subcycle timescales due to resonances and excitation (related to multiphoton absorption in quantum mechanics~\cite{Barth2011,Barth2014}). 
A subcycle adiabatic approximation is usually made when the ionization process occurs faster than the variations of the laser field, i.e., on a timescale sufficiently smaller than one laser cycle. In this approximation, the laser field is considered frozen in time, i.e., $\mathbf{E}(t) = \mathbf{E}$. An effective potential, called LF-effective potential in what follows, is given by 
\begin{equation}
    \label{eq:Veff_LF}
    V_{{\rm LF},k}(\mathbf{r},t) = V_k( \mathbf{r}) + \mathbf{r} \cdot \mathbf{E}.
\end{equation}
This effective potential is depicted on the left panel of Fig.~\ref{fig:Veff_LF_RF}. In the adiabatic approximation in the LF, there are two fixed points for $|\mathbf{E}|^2\leq 4Z_k/(27 a^4)$: The stable point at the bottom of the potential well (magenta dot in the right panel of Fig.~\ref{fig:Veff_LF_RF}) and the saddle point at the top of the potential barrier induced by the laser field (green dot in the right panel of Fig.~\ref{fig:Veff_LF_RF}). At $|\mathbf{E}|^2= 4Z_k/(27 a^4)$, the stable point and the saddle point merge at $r=a/\sqrt{2}$. For larger effective intensities, the potential has no fixed points.
\par
The location of the fixed points is at ${\bf r} = \mathbf{r}_{\alpha,k}$ and $\mathbf{p}_{\alpha,k} = \boldsymbol{0}$, where $\alpha = {\rm b}$ for the stable point at the bottom of the potential well, and $\alpha = {\rm s}$ for the stable point at the top of the potential barrier. The position of each fixed point is solution of
$$
{\bm \nabla} V_k ( \mathbf{r}_{\alpha,k} ) + \mathbf{E} = \boldsymbol{0} ,
$$
which leads to $\mathbf{r}_{\alpha,k} = - r_{\alpha,k} {\bf E}/\vert {\bf E}\vert$ and $r_{\alpha,k}$ solution of 
$$
-Z_k r_{\alpha,k}/(r_{\alpha,k}^2+a^2)^{3/2}+ E_0 f(t) = 0.
$$
The location of these adiabatic fixed points changes on a timescale smaller than the period of the laser field, since $\mathbf{r}_{\alpha,k}$ rotates with the laser field. We notice that the radii $r_{\alpha,k}$ do not change on a subcycle timescale. 

\subsubsection{Potential well \label{sec:maximum_intensity}}
The stable point at the bottom of the potential well plays an important role for the double ionization mechanisms: It determines the intensity at which the double ionization saturates in the SDI regime, and it bounds the inner electron during the excursion of the outer electron in the NSDI regime~\cite{Mauger2009}. 
\par 
After the ionization of the first electron $\mathrm{e}_1^-$, electron $\mathrm{e}_2^-$ is near the bottom of the potential well. If there exists a time $t$ at which the stable point does no longer exists, $\mathrm{e}_2^-$ is also no longer bounded, and as a consequence ionizes. The lowest laser intensity at which there exists a time $t$ such that $|\mathbf{E}(t)|^2 \geq 4Z_k/(27 a^4)$ is for $E_0^2 \geq 4Z_k/(27 a^4)$. At intensity
\begin{equation}
\label{eq:bounded_region_Imax_approximation_LF}
I_{\max} \approx \dfrac{8Z_2^2}{27 a^4} ,
\end{equation}
there are no bound regions in phase space near the bottom of the well for the inner electron to remain bounded during the plateau. 
We observe that Eq.~\eqref{eq:bounded_region_Imax_approximation_LF} is given by the same formula as the linearly polarized (LP) case with the adiabatic approximation~\citep{Mauger2009}. This is consistent with the adiabatic approximation since only the peak amplitude of the laser is important. The value of the maximum intensity in the adiabatic approximation in the LF is  $I_{\max}=5.1\times 10^{14} \;\mathrm{W}\cdot \mathrm{cm}^{-2}$ for Mg ($a=3$), and $I_{\max}=8.2 \times 10^{15} \; \mathrm{W}\cdot\mathrm{cm}^{-2}$ for Ar. 

\subsubsection{Potential barrier}
In the adiabatic approximation in the LF, the saddle point exists for $(E_0 f(t))^2 < I_{\max}/2$. Near the saddle point, the LF-effective potential reads
$$
V_{{\rm LF},k} ( - r \hat{\mathbf{x}} ) = h_{{\rm s},k} (t) \left[ 1 + \dfrac{(r - r_{{\rm s},k} )^2}{2 W_{{\rm s},k} (t)^2} \right] + O ( \left( r-r_{{\rm s},k} \right)^3) ,
$$
where $h_{{\rm s},k} (t)$ and $W_{{\rm s},k} (t)$ are the height and the width of the potential barrier in the LF, respectively. The energy of the saddle point is given by $h_{{\rm s},k}(t) = H_k ( \mathbf{r}_{{\rm s},k} , \boldsymbol{0},t)=-E_0 f(t) (2r_{{\rm s},k}^2+a^2)/r_{{\rm s},k}$ which also does not change on a subcycle timescale. Its width for the soft-Coulomb potential is given by $W_{{\rm s},k}(t)= [ (r_{{\rm s},k}^2+a^2)(2r_{{\rm s},k}^2+a^2)/(2r_{{\rm s},k}^2-a^2) ]^{1/2}$.
\par
A good approximation is to assume that this saddle point is rather far away from the core, i.e., $r_{{\rm s},k}\gg a$. In this region, the soft-Coulomb potential resembles a hard-Coulomb one, $ - Z_k / |\mathbf{r} |$. Under this approximation, the location of the saddle point, its energy and the width of the barrier are  
\begin{subequations}
\begin{eqnarray}
\label{eq:rskLF}
&& r_{{\rm s},k} = \sqrt{ Z_k / |\mathbf{E}(t)| } , \\
\label{eq:hskLF}
&& h_{{\rm s},k}(t) = - 2 \sqrt{Z_k |\mathbf{E}(t)|}, \\
\label{eq:WskLF}
&& W_{{\rm s},k} (t) = \sqrt{ Z_k/|\mathbf{E}(t)| } .
\end{eqnarray}
\end{subequations}
\par
Performing the adiabatic approximation in the LF suppresses all dynamical content beyond the subcycle timescale even if the distance between the saddle and the core, the energy and the width of the potential barrier do not change on a subcycle timescale. This adiabatic approximation should be used with extra caution, depending on the parameters of the laser and the atom. 

\subsection{Adiabatic approximation in the RF}

\begin{figure}
	\centering
	\includegraphics[width=.5\textwidth]{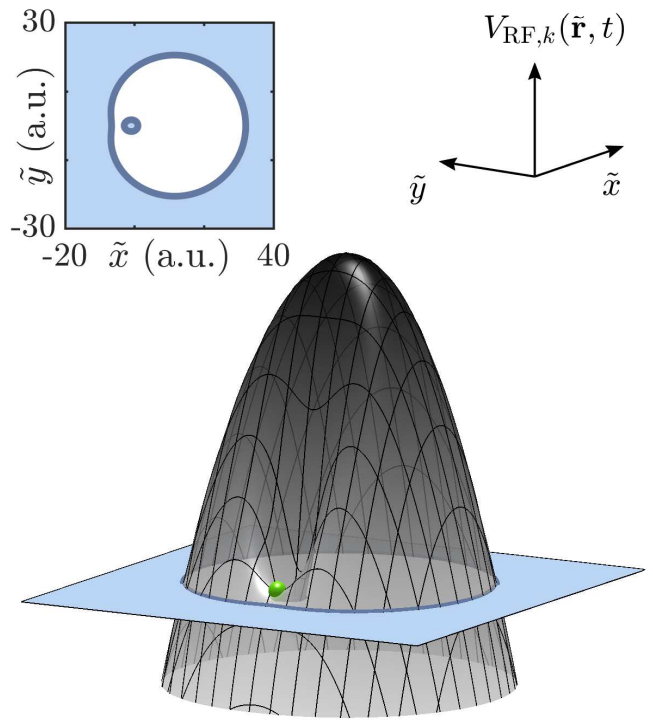}
	\caption{RF-effective potential $V_{{\rm RF},k} (\tilde{\mathbf{r}},t)$ given by Eq.~\eqref{eq:zero_velocity_surface} for $I = 10^{14} \; \mathrm{W} \cdot \mathrm{cm}^{-2}$, $f=1$, $a=3$ and $k=2$. The green ball indicates the location of the saddle point. The section corresponds to the surface $\tilde{H}_k ( \tilde{\mathbf{r}} , \tilde{\mathbf{p}} , t) = -0.6$ in the plane $(\tilde{x},\tilde{y})$. The blue colored areas correspond to the allowed positions of the electron.}
	\label{fig:zero_velocity_surface}
\end{figure}

Hamiltonians~\eqref{eq:Hamiltonian_RF_1ae} can be rewritten as
\begin{eqnarray*}
\tilde{H}_k ( \tilde{\mathbf{r}} , \tilde{\mathbf{p}} , t) =&& \dfrac{1}{2} \left| \tilde{\mathbf{p}} + \omega \tilde{\mathbf{r}} \times \hat{{\mathbf{z}}} \right|^2 \nonumber \\
&&- \dfrac{\omega^2}{2} | \tilde{\mathbf{r}} |^2 + V_k ( \tilde{\mathbf{r}} ) + \tilde{\mathbf{r}} \cdot \tilde{\mathbf{E}} (t) ,
\end{eqnarray*}
i.e., a kinetic energy plus an effective potential, called the RF-effective potential in what follows, given by 
\begin{equation}
\label{eq:zero_velocity_surface}
V_{{\rm RF},k} ( \tilde{\mathbf{r}} , t) = - \dfrac{\omega^2}{2} |\tilde{\mathbf{r}} |^2 + V_k ( \tilde{\mathbf{r}} ) + \tilde{\mathbf{r}} \cdot \tilde{\mathbf{E}}(t) .
\end{equation}
We observe that, for all times, the phase-space coordinates of the $k$-th electron is such that $\tilde{H}_k ( \tilde{\mathbf{r}} , \tilde{\mathbf{p}} , t) \geq V_{{\rm RF},k} ( \tilde{\mathbf{r}} , t)$. The condition $\tilde{H}_k ( \tilde{\mathbf{r}} , \tilde{\mathbf{p}}, t) = V_{{\rm RF},k} ( \tilde{\mathbf{r}}, t)$ is fulfilled if and only if the velocity of the electron in the RF vanishes, i.e., $\dot{\tilde{\mathbf{r}}} = \tilde{\mathbf{p}} + \omega \tilde{\mathbf{r}} \times \hat{{\mathbf{z}}} = \boldsymbol{0}$. The surface $\tilde{H}_k ( \tilde{\mathbf{r}}, \tilde{\mathbf{p}},t) = V_{{\rm RF},k} ( \tilde{\mathbf{r}}, t)$ is usually called the zero-velocity surface in the literature~\cite{Hill1878, Mauger2010_PRL}.
\par
Figure~\ref{fig:zero_velocity_surface} shows the RF-effective potential for $I = 10^{14} \; \mathrm{W}\cdot \mathrm{cm}^{-2}$, $a = 3$ and $f=1$ in the two-dimensional case. A slice of this RF-effective potential is also shown on the right panel of Fig.~\ref{fig:Veff_LF_RF}. The RF-effective potential is composed of a well near the origin due to the ion-electron interaction $V_k (\tilde{\mathbf{r}})$. The dome shape of the RF-effective potential comes from the Coriolis force. 
Furthermore, we notice that the Coriolis term in the RF-effective potential lowers the effective potential compared with the LF-effective potential (in particular, it lowers the potential barrier), i.e.,
\begin{equation}
    \label{eq:Veff_RF}
    V_{{\rm RF},k}({\bf r},t)< V_{{\rm LF},k}({\bf r},t).
\end{equation}
For $f=1$, the energy $\tilde{H}_k ( \tilde{\mathbf{r}}, \tilde{\mathbf{p}},t)$ does no longer depend explicitly on time, and as a consequence, it is a constant, known as the Jacobi constant~\citep{Jacobi1836} $\mathcal{K} = \tilde{H}_k( \tilde{\mathbf{r}}, \tilde{\mathbf{p}})$ in celestial mechanics. Notice that in contrast, the energy $H_k ( \mathbf{r}, \mathbf{p},t)$ in the LF is not conserved, even for $f=1$. If the electron is inside the well near the origin and $\mathcal{K}$ is smaller than the energy of the saddle point (see the right panel of Fig.~\ref{fig:Veff_LF_RF}), the electron is topologically bounded by the RF-effective potential, as observed in the inset of Fig.~\ref{fig:zero_velocity_surface}. In this case, the electron is trapped and cannot ionize classically. The same picture holds for time-dependent Hamiltonian in the adiabatic approximation in the RF, which is accurate as long as $\tilde{H}_k ( \tilde{\mathbf{r}} , \tilde{\mathbf{p}} , t)$ varies slowly with respect to time, i.e., if the ramp-up duration is not smaller than one or two laser periods. 
\par
Next, we perform an adiabatic approximation on a timescale of the variations of the laser envelope, i.e., assuming $f (t)$ is constant, and not ${\bf E}(t)$ constant as in the previous adiabatic approximation in the LF. In this article, we refer to the nonadiabatic effects as the effects which occur on subcycle time scales, i.e., which are included in the adiabatic approximation in the RF and not included in the adiabatic approximation in the LF. The adiabatic approximation in the RF gives rise to the existence of three fixed points: The stable point at the bottom of the well, the saddle point at the top of the potential barrier, and the fixed point at the top of the RF-effective potential (see Figs.~\ref{fig:Veff_LF_RF} and~\ref{fig:zero_velocity_surface}). The location of the fixed points is at $\mathbf{r} = \mathbf{r}_{\alpha,k}$ and $\tilde{\mathbf{p}}_{\alpha,k} = \omega \hat{{\mathbf{z}}} \times \tilde{\mathbf{r}}_{\alpha,k}$, where $\alpha = {\rm b}$ for the stable point at the bottom of the potential well and $\alpha = {\rm s}$ for the saddle point at the top of the potential barrier. The position of the fixed points is solution of
\begin{equation}
\label{eq:fixed_point_definition}
- \omega^2\tilde{\mathbf{r}}_{\alpha,k} + {\bm \nabla} V_k ( \tilde{\mathbf{r}}_{\alpha,k} ) + \tilde{\mathbf{E}} = \boldsymbol{0} ,
\end{equation}
which leads to $\tilde{\mathbf{r}}_{\alpha,k} = -\tilde{r}_{\alpha,k} \tilde{\bf E}/\vert {\bf E} \vert$, where $\tilde{r}_{\alpha,k}$ is a solution of
$$
\omega^2 \tilde{r}_{\alpha,k} -Z_k \tilde{r}_{\alpha,k}/(\tilde{r}_{\alpha,k}^2+a^2)^{3/2} + E_0 f(t) = 0 .
$$

\begin{figure}
	\centering
	\includegraphics[width=.5\textwidth]{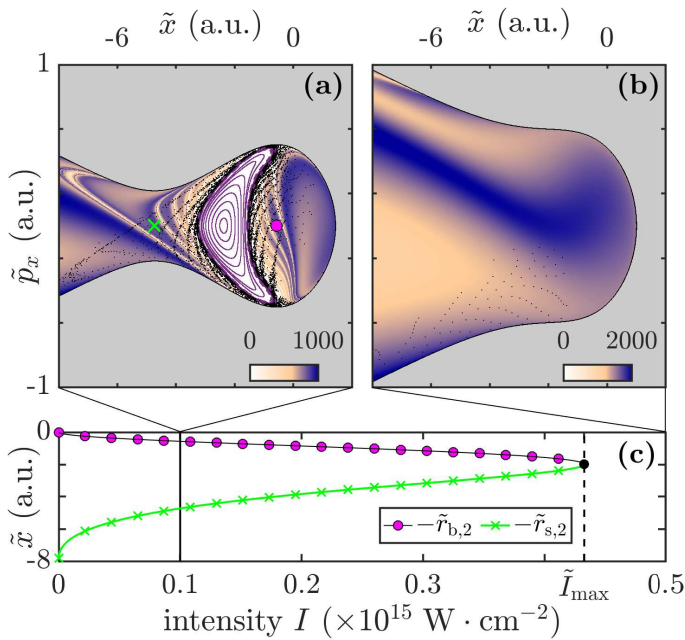}
	\caption{(a--b) Poincar\'{e} sections of Hamiltonian~\eqref{eq:Hamiltonian_RF_1ae} for $\tilde{y} = 0$ and $\dot{\tilde{y}} > 0$ in the plane $(\tilde{x}, \tilde{p}_{x})$ for $\mathrm{Mg}$ ($a=3$), $k=2$ (inner electron) and $f=1$. Trajectories with quasi-periodic motions are in purple. The color scale is the final distance of the electron from the core with initial condition given by $(\tilde{x}, \tilde{p}_{x})$ integrated over ten laser cycles. The energy surface is $\tilde{H}_2 ( \tilde{\mathbf{r}} , \tilde{\mathbf{p}}) = - 0.55$. The gray surfaces are the classical forbidden regions. The intensity is (a) $I = 10^{14} \; \mathrm{W}\cdot\mathrm{cm}^{-2}$ and (b) $I = 5\times 10^{14} \; \mathrm{W}\cdot\mathrm{cm}^{-2}$. (c) Position along $\hat{\tilde{\mathbf{x}}}$ of the fixed points: the saddle point at $\tilde{x} = -\tilde{r}_{{\rm s},2}$ (green curve) and the bottom of the well near the origin at $\tilde{x} = - \tilde{r}_{{\rm b},2}$ as a function of the laser intensity. In panel (a), the fixed points are indicated by a green cross (at $\tilde{x} = - \tilde{r}_{{\rm s}, 2}$) and a magenta dot (at $\tilde{x} = - \tilde{r}_{{\rm b},2}$).}
	\label{fig:poincare_section}
\end{figure}

Since the adiabatic approximation in the RF allows us to look at the dynamics beyond the subcycle timescale, we investigate the dynamics of the inner electron, modeled by Hamiltonian~\eqref{eq:Hamiltonian_RF_1ae} for $k=2$ and $f=1$.
Figure~\ref{fig:poincare_section} shows a Poincar\'{e} section of Hamiltonian~\eqref{eq:Hamiltonian_RF_1ae} for $\tilde{y} = 0$ and $\dot{\tilde{y}} > 0$ and an energy close to second ionization potential of $\mathrm{Mg}$. At intensity $I = 10^{14} \; \mathrm{W}\cdot\mathrm{cm}^{-2}$, we observe that the electron on this energy surface is not topologically bounded and can ionize over the barrier. However, there are invariant tori near the origin in phase space which prevent ionization. On these tori, the motion of the electron is quasi-periodic, and the electron is dynamically bounded. The topological condition [$\tilde{H}_k(\tilde{\mathbf{r}}_k,\tilde{\mathbf{p}}_k,t) > \tilde{h}_{{\rm s},k} (t)$, where $\tilde{h}_{{\rm s},k}(t)$ denotes the height of the potential barrier] is a necessary condition for ionizing, but not a sufficient one. In other words, there are two types of obstructions for ionization: topological ones where the allowed region is bounded by the RF-effective potential, and dynamical ones where the electron could be sitting on invariant tori.  

\subsubsection{Potential well}

In the RF, the stable fixed point is located at $\tilde{\bf r}_{{\rm b},k}=-\tilde{r}_{{\rm b},k} \tilde{\bf E}/\vert {\bf E}\vert$ where $\tilde{r}_{{\rm b},k}$ is a solution of Eq.~\eqref{eq:fixed_point_definition}. 
The invariant tori reflect the existence and the stability of the fixed point at the bottom of the well near the origin. Figure~\ref{fig:poincare_section}c shows the position of the saddle point $\tilde{r}_{{\rm s},2}$ and of the bottom of the well $\tilde{r}_{{\rm b},2}$ as a function of the laser intensity.  At $I = \tilde{I}_{\max}$, the saddle point and the stable fixed point merge. For this value of intensity, the stable fixed point is solution of $\partial V_{{\rm RF},2} ( -\tilde{r} \hat{\tilde{\mathbf{x}}} ) / \partial \tilde{r} = 0$ [corresponding to Eq.~\eqref{eq:fixed_point_definition}] and $\partial^2 V_{{\rm RF},2} ( -\tilde{r} \hat{\tilde{\mathbf{x}}} ) / \partial \tilde{r}^2 = 0$. From these two conditions, we obtain 
\begin{subequations}
\label{eq:exact_Imax_definition}
\begin{eqnarray}
\label{eq:bounded_region_Imax_complete_E0_expression}
\tilde{I}_{\max} &=& \dfrac{18 Z_2^2 \tilde{r}_{{\rm b},2}^6}{\left( \tilde{r}_{{\rm b},2}^2 + a^2 \right)^{5}} , \\
\label{eq:bounded_region_Imax_complete}
\dfrac{\omega^2}{Z_2} &=& \dfrac{a^2 - 2 \tilde{r}_{{\rm b},2}^2}{\left( \tilde{r}_{{\rm b},2}^2 + a^2 \right)^{5/2}} .
\end{eqnarray}
\end{subequations}
Considering that $\omega^2/Z_2$ is very small (compared to $1/a^3$), we obtain $\tilde{r}_{{\rm b},2} \approx  a/\sqrt{2} - (9 \sqrt{3}/16) a^4 \omega^2/Z_2$. This approximation is substituted in Eq.~\eqref{eq:bounded_region_Imax_complete_E0_expression} to get
\begin{equation}
\label{eq:bounded_region_Imax_approximation}
\tilde{I}_{\max} = I_{\max} \left( 1 - \omega^2 \dfrac{9 a^3}{\sqrt{6} Z_2} \right) + O \left( \dfrac{a^6 \omega^4}{Z^2} \right) ,
\end{equation}
where $I_{\max}$ is the threshold intensity in the adiabatic approximation in the LF [see Eq.~\eqref{eq:bounded_region_Imax_approximation_LF}]. For a laser wavelength $\lambda=780 \; \mathrm{nm}$, $\tilde{I}_{\max} = 4.3\times 10^{14} \; \mathrm{W} \cdot \mathrm{cm}^{-2}$ for Mg ($a=3$) instead of $I_{\max} = 5.1 \times 10^{14} \; \mathrm{W} \cdot \mathrm{cm}^{-2}$ in the adiabatic approximation in the LF. For Ar ($a=1.5$), $\tilde{I}_{\max} = 8.0 \times 10^{15} \; \mathrm{W} \cdot \mathrm{cm}^{-2}$ instead of $I_{\max} = 8.2 \times 10^{15} \; \mathrm{W} \cdot \mathrm{cm}^{-2}$. These values are in agreement with Figs.~\ref{fig:probability_curves} and \ref{fig:probability_curves_Ar}. We observe that the leading term in Eq.~\eqref{eq:bounded_region_Imax_approximation} is the same as for LP case with the adiabatic approximation~\citep{Mauger2009} [see also Eq.~\eqref{eq:bounded_region_Imax_approximation_LF}]. The difference here, is that the energy gained by the electron in the LF on subcycle timescales is taken into account in the adiabatic approximation in the RF, and we observe a correction which depends on the laser frequency as $\omega^2$. For $I > \tilde{I}_{\max}$, after the ionization of the first electron and at the peak of the laser amplitude, we do not observe any invariant tori in Fig.~\ref{fig:poincare_section}b, i.e., there is no region in phase space where the electron remains bounded. 

\subsubsection{Potential barrier}
The saddle point of the RF-effective potential, represented by a green ball in Fig.~\ref{fig:zero_velocity_surface}, is crucial for ionization. In particular, it controls the features of the ionization barrier, its location, its energy and its width. It exists for $(E_0 f(t))^2 < \tilde{I}_{\max}/2$. From the location of the saddle point $\tilde{r}_{{\rm s},k}$ solution of Eq.~\eqref{eq:fixed_point_definition}, we deduce its energy
\begin{equation}
\label{eq:hRFk}
    \tilde{h}_{{\rm s}, k}(t)=-\frac{\omega^2}{2}\tilde{r}_{{\rm s},k}^2+V_k(-\tilde{r}_{{\rm s},k}\hat{\tilde{{\bf x}}})-\tilde{r}_{{\rm s},k} E_0 f(t).
\end{equation}
In order to get some analytical insights into the potential barrier in the RF, we distinguish two regimes of intensity: low and high effective intensities $(E_0 f(t))^2$. 
\par
For high effective intensities and for a hard-Coulomb potential, we approximate the solution of Eq.~\eqref{eq:fixed_point_definition} by
$$
\tilde{r}_{{\rm s},k} (t) = r_{{\rm s},k} (t) \left[ 1 - \dfrac{\omega^2\sqrt{Z_k}}{2 |\mathbf{E}(t)|^{3/2}}\right] + O\left( \dfrac{1}{|\mathbf{E}(t)|^3} \right) ,
$$
where $r_{{\rm s},k} (t)$ is the radius of the saddle point in the adiabatic approximation in the LF [see Eq.~\eqref{eq:rskLF}]. In the RF, the saddle point is closer to the origin. We observe that $|\mathbf{E}(t)|/(Z_k \omega^4)^{1/3}$, which depends on the effective amplitude of the laser field $|\mathbf{E}(t)|$, governs the location of the saddle point in the adiabtic approximation in the LF compared to its location in the adiabatic approximation in the RF. The height and the width of the potential barrier in the RF are given by
\begin{eqnarray*}
&& \tilde{h}_{{\rm s},k} (t) = h_{{\rm s},k} (t) \left[ 1 + \dfrac{\omega^2 \sqrt{Z_k}}{4 |\mathbf{E}(t)|^{3/2}}\right] + O\left( \dfrac{1}{|\mathbf{E}(t)|^3} \right)  , \\
&& \tilde{W}_{{\rm s},k} (t) = W_{{\rm s},k} (t) \left[ 1 - \dfrac{7 \omega^2 \sqrt{Z_k}}{8 |\mathbf{E}(t)|^{3/2}}\right] + O\left( \dfrac{1}{|\mathbf{E}(t)|^3} \right)  ,
\end{eqnarray*}
where $h_{{\rm s},k} (t)$ and $W_{{\rm s},k}(t)$ are the height and the width of the potential barrier in the adiabatic approximation in the LF [see Eqs.~\eqref{eq:hskLF} and~\eqref{eq:WskLF}], respectively. In the RF, the height of the saddle point is lower and its width is smaller. The decrease of the height of the barrier results from the energy gained by the electrons on subcycle timescales in the LF due to resonances and excitations, which are naturally embedded in the adiabatic approximation in the RF. As a consequence, over-the-barrier ionization happens at lower laser intensities in the adiabatic approximation in the RF than it happens in the adiabatic approximation in the LF. This is studied in Sec.~\ref{sec:SDI_mechanisms}. In addition, due to the decrease of the width of the barrier in the adiabatic approximation in the RF, the electron would tunnel-ionize closer to the core than in the adiabatic approximation in the LF. This is also what is observed in the tunnel-ionization theory where the Coulomb potential is treated as a perturbation of the laser interaction~\cite{PerelomovI1966, PerelomovII1967, PerelomovIII1967}. 
\par
For low effective intensities and for a hard-Coulomb potential, we approximate the solution of Eq.~\eqref{eq:fixed_point_definition} by
$$
\tilde{r}_{{\rm s},k}(t)=\left(\frac{Z_k}{\omega^2}\right)^{1/3} \left[ 1 - \frac{|\mathbf{E}(t)|}{3 (Z_k \omega^4 )^{1/3}}\right]+O\left( |\mathbf{E}(t)|^2\right)  .
$$
In contrast to the LF-effective potential, a saddle point exists also when the effective amplitude of the laser field is $E_0 f(t) = 0$ [see Eq.~\eqref{eq:rskLF} for comparison]. The height and the width of the potential barrier are approximately given by
\begin{subequations}
\begin{eqnarray}
\label{eq:hRFkapp}
&& \tilde{h}_{{\rm s},k}(t)=-\frac{3}{2}(Z_k\omega)^{2/3} \left[ 1 + \dfrac{2 |\mathbf{E}(t)|}{3 (Z_k \omega^4)^{1/3}}\right] \nonumber \\
&& \qquad \qquad \qquad \qquad \qquad \qquad +O \left( |\mathbf{E}(t)|^2 \right)  , \\
&& \tilde{W}_{{\rm s},k}(t)=\dfrac{1}{\sqrt{2}} \left( \dfrac{Z_k}{\omega^2} \right)^{1/3} + O \left( |\mathbf{E}(t)|^2 \right) , 
\end{eqnarray}
\end{subequations}
respectively. We notice the weak dependence of the width with respect to the intensity of the field. At low effective intensities, most of the ionizations occur due to resonances and excitations, i.e., multiphoton absorption in the quantum mechanical treatment of ionization. Contrary to the high effective intensity regime, the importance of the intensity in triggering ionization is drastically decreased. The frequency of the laser is the main parameter for ionization. 
\par
In this low effective intensity regime, the main parameter is the laser frequency, whereas in the high effective intensity regime, the main parameter is the intensity of the laser. In the hard-Coulomb approximation, Eq.~\eqref{eq:fixed_point_definition} becomes a cubic polynomial. Its discriminant vanishes at $|\mathbf{E}(t)|=3(Z_k\omega^4/4)^{1/3}$, and the solution jumps from the low-effective intensity branch to the high-effective intensity branch. For $Z_k=1$ and $\omega=0.0584$, this effective intensity is around $10^{14}\ \mathrm{W}\cdot\mathrm{cm}^{-2}$.

\section{Ionization times for the sequential double ionization \label{sec:SDI_mechanisms}}

\begin{figure*}
	\centering
	\includegraphics[width=\textwidth]{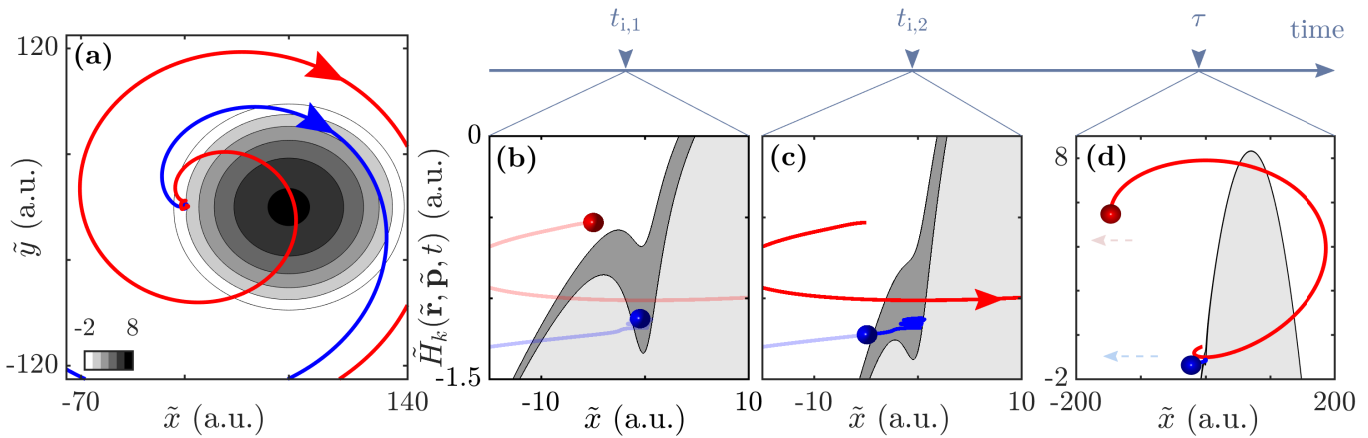}
	\caption{Typical SDI of Hamiltonian~\eqref{eq:Hamiltonian_NSDI_LF} in the RF for Ar ($\mathcal{E}_g=-1.60$, $a=1.5$), $I = 5\times 10^{14} \unites{W}\cdot \mathrm{cm}^{-2}$ and a trapezoidal envelope 2--4--2 (duration of ramp-up $\tau = 2 T$). The red and blue curves are the outer (first ionized) and inner (second ionized) electron trajectories for $t \in [t_{{\rm i},1} , 6 T]$, respectively. (a) Projection in the plane $(\tilde{x},\tilde{y})$. A contour plot of the RF-effective potential $V_{{\rm RF},1} ( \tilde{\mathbf{r}}_1,\tilde{\mathbf{p}}_1 ,\tau)$ is displayed on the leftmost panel. (b--d) Snapshots of the electrons in the plane $( \tilde{x} , \tilde{H}_k ( \tilde{\mathbf{r}}, \tilde{\mathbf{p}},t))$. The dark and light gray areas are the classical forbidden regions delimited by the RF-effective potential of the outer electron $V_{{\rm RF},1} ( \tilde{x} \hat{\tilde{\mathbf{x}}},t)$ and of the inner electron $V_{{\rm RF},2} ( \tilde{x} \hat{\tilde{\mathbf{x}}},t)$, respectively. The red and blue dots are the outer and inner electrons at times (b) $t_{{\rm i},1} \approx  0.8 T$, (c) $t_{{\rm i},2} \approx 1.9 T$, (d) $\tau = 2T$. (d) The light red and blue dashed arrows indicate the subsequent ionization of the electrons.}
	\label{fig:scenario_SDI}
\end{figure*}

The simplest mechanism for double ionization is that each electron undergoes an ionization process sequentially. In this section, we analyze the sequential double ionization process, and in particular the ionization times of the two electrons, using the SAE models given by Hamiltonians~\eqref{eq:Hamiltonian_RF_1ae}. 
\par
Figure~\ref{fig:scenario_SDI} shows a typical SDI of Hamiltonian~\eqref{eq:Hamiltonian_NSDI_LF} depicted in the RF. Notice that the dynamics of the electrons is the same in the LF than in the RF, i.e., all nonadiabatic effects are included in both Hamiltonian~\eqref{eq:Hamiltonian_NSDI_LF} and Hamiltonian~\eqref{eq:Hamiltonian_NSDI_RF}. Here, we describe the ionization processes of a typical SDI in the RF, but the same observations would hold in the LF. Nonadiabatic effects are revealed when we compare the physical observables obtained in the adiabatic approximation in the LF, with the ones obtained in the adiabatic approximation in the RF, since the latter naturally embeds the energy gained by the electron on subcycle timescales in the LF. Figure~\ref{fig:scenario_SDI} shows a typical SDI in the RF. In Figs.~\ref{fig:scenario_SDI}b--d, the light and dark gray areas are the classical forbidden regions delimited by the RF-effective potential of the outer electron $V_{{\rm RF},1} ( \tilde{x} \hat{\tilde{\mathbf{x}}},t)$ and of the inner electron $V_{{\rm RF},2} ( \tilde{x} \hat{\tilde{\mathbf{x}}},t)$, respectively. Initially, both electrons are topologically bounded by the RF-effective potential and cannot ionize. During the ramp-up, when time increases, the height of the potential barrier decreases [from Eq.~\eqref{eq:fixed_point_definition}, one has $\dot{\tilde{h}}_{{\rm s},k} (t) = -\tilde{r}_{{\rm s},k} (t) E_0 \dot{f}(t)$]. The first electron ionizes at time $t_{{\rm i},1}$, when its energy is above the saddle point of its RF-effective potential, i.e., $\tilde{H}_1 (\tilde{\mathbf{r}} , \tilde{\mathbf{p}} , t_{{\rm i},1}) \approx \tilde{h}_{{\rm s},1} (t_{{\rm i},1})$. At this time, the second electron remains bounded by its RF-effective potential. The second electron ionizes at time $t_{{\rm i},2}$, when its energy is above the saddle point of its RF-effective potential, i.e., $\tilde{H}_2 (\tilde{\mathbf{r}} , \tilde{\mathbf{p}} , t_{{\rm i},2}) \approx \tilde{h}_{{\rm s},2} (t_{{\rm i},2})$. Hence, the saddle point plays a central role for the conditions under which the electron ionizes. 
\par
Initially, the two-active electron atom is in the ground state of energy $\mathcal{E}_g = \mathcal{E}_1 + \mathcal{E}_2$, where $\mathcal{E}_k$ is the initial energy of the $k$-th electron. 
Following Ref.~\citep{Wang2012,Lan2014}, we assume that the electrons ionize independently and that the initial energy of the inner (resp.\ outer) electron corresponds to the first (resp.\ second) ionization potential of the atom. This assumption allowed for the agreement of the ionization times obtained from purely classical models in the inner-outer approximation with $H_k ( \mathbf{r} , \mathbf{p} ,0) = \mathcal{E}_k$, with the ionization times obtained experimentally~\citep{Pfeiffer2011,Wang2012,Lan2014}.

\subsection{Adiabatic approximation in the LF\label{eq:ionization_time_adiabatic}}

First, we perform the adiabatic approximation in the LF, i.e., we consider that the energy is conserved in the LF before ionization, i.e., $H_k ( \mathbf{r} , \mathbf{p} ,t) = \mathcal{E}_k$. The top of the effective potential is at energy $h_{{\rm s},k} ( t )=-2(Z_k E_0 f(t))^{1/2}$. The time at which the $k$-th electron can ionize over the barrier corresponds to the time at which its energy $\mathcal{E}_k$ gets over the top of the effective potential, i.e., $\mathcal{E}_k \approx h_{{\rm s},k} (t)$. This condition on the ionization time $t_{{\rm i},k}$ is given by 
\begin{equation}
\label{eq:effective_laser_amplitude_adiabatic}
E_k^{\star} = \dfrac{|\mathcal{E}_k|^2}{4 Z_k} ,
\end{equation}
where $E_k^{\star} = {E}_0 f(t_{{\rm i},k})$ is the effective amplitude of the laser field at ionization of the electron $\mathrm{e}_k^-$. In terms of intensity, the minimum intensity for which ionization occurs over the barrier is obtained when $f(t)=1$, and it is given by 
\begin{equation}
\label{eq:ImaxLF}
I_{\rm OTB , k} = \dfrac{|\mathcal{E}_k|^4}{8 Z_k^2}.
\end{equation}
This threshold intensity depends on the effective charge of the ion and the ionization potential of the electron $\mathcal{E}_k$ only. In particular, it does not depend on the laser frequency. For $\mathrm{Mg}$, its value is $I_{\rm OTB,1} \approx 2.7 \times 10^{13} \; \mathrm{W}\cdot \mathrm{cm}^{-2}$, and $I_{\rm OTB,1} \approx 5.7 \times 10^{14} \; \mathrm{W}\cdot \mathrm{cm}^{-2}$ for Ar.

\subsection{Adiabatic approximation in the RF}

\begin{figure}
	\centering
	\includegraphics[width=.5\textwidth]{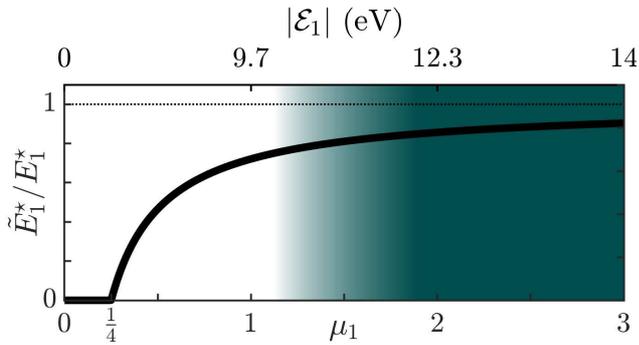}
	\caption{Ratio of the effective amplitude of the laser field in the nonadiabatic $\tilde{E}_k^{\star}$ [see Eq.~\eqref{eq:effective_amplitude}] over the adiabatic $E_k^{\star}$ [see Eq.~\eqref{eq:effective_laser_amplitude_adiabatic}] for the first ionized electron ($k=1$) as a function of the parameter $\mu_k = 2 |\mathcal{E}_k|^3 (Z_k \omega)^{-2}/27$. The upper axis indicates the ionization potential of the first electron $|\mathcal{E}_1|$ for $\omega = 0.0584$ and $Z_1 = 1$. The shaded region indicates also the ratio $\tilde{E}_1^{\star}/E_1^{\star}$: The dark region is where the adiabatic approximation is similar to the nonadiabatic case, and the light region is where the adiabatic approximation fails to predict the ionization process of the electron. }
	\label{fig:adiabatic_vs_nonadiabatic}
\end{figure}

In the RF, we consider $\tilde{H}_k ( \tilde{\mathbf{r}} , \tilde{\mathbf{p}} ,t) = \mathcal{E}_k$. 
For $t < t_{{\rm i},1}$, both electrons are topologically bounded by their respective RF-effective potential and/or dynamically bounded. For increasing $f(t)$, the saddle point of $V_{{\rm RF},1}$ goes down. At time $t_{{\rm i},1}$, the energy of the outer electron $\mathrm{e}_1^-$ reaches the level of the saddle point of $V_{{\rm RF},1}$, i.e., $\tilde{h}_{{\rm s},1} (t_{{\rm i},1}) = \mathcal{E}_1$, and ionizes over the barrier subsequently, while the energy of inner electron $\mathrm{e}_2^-$ is still lower than the one of the saddle point of $V_{{\rm RF},2}$. Later, at time $t_{{\rm i},2}$, the energy of $\mathrm{e}_2^-$ reaches the level of the saddle point of $V_{{\rm RF},2}$, i.e., $\tilde{h}_{{\rm s},2} (t_{{\rm i},2}) = \mathcal{E}_2$, and ionizes above the barrier.
\par
Combining Eq.~\eqref{eq:fixed_point_definition} and the condition $V_{{\rm RF},k} (\tilde{\bf r}_{{\rm s},k},t_{{\rm i},k}) = \mathcal{E}_k$, we obtain the two conditions
\begin{subequations}
\label{eq:saddle_point_ionization_RF}
\begin{eqnarray}
&& {\cal E}_k=V_{{\rm RF},k} (-\tilde{r}_{{\rm s},k} \hat{\tilde{{\bf x}}}, t_{{\rm i},k}),\\
&& {\bm \nabla} V_{{\rm RF},k}(-\tilde{r}_{{\rm s},k} \hat{\tilde{{\bf x}}}, t_{{\rm i},k})= {\bf 0}.
\end{eqnarray}
\end{subequations}
Close to the saddle point, we approximate the ion-electron potential by a hard-Coulomb potential $V_k ( \tilde{\mathbf{r}} ) = -Z_k / |\tilde{\mathbf{r}}|$. Equations~\eqref{eq:saddle_point_ionization_RF} does not have a solution with positive $\tilde{r}_{{\rm s},k}$ for $\mu_k < 1/4$, where 
$$
\mu_k = \frac{2 |\mathcal{E}_k|^3}{ 27 (Z_k \omega)^{2}} .
$$
The parameter $\mu_k$ is dimensionless, and measures the magnitude of the nonadiabatic effects resulting from the electron energy variations on subcycle timescales happening in the LF, as shown below for the threshold intensity at which over-the-barrier ionization can occur and the ionization time of the electrons. If $\mu_k < 1/4$, the electron is not topologically bounded by the RF-effective potential before the laser field is turned on, i.e., ${\cal E}_k > \tilde{h}_{{\rm s},k} (0)$ with $\tilde{h}_{{\rm s},k}(0) = -3 (Z_k \omega)^{2/3}/2$ [see also Eq.~\eqref{eq:hRFkapp}] (meanwhile the electron is dynamically bounded). Consequently, Eqs.~\eqref{eq:saddle_point_ionization_RF} do not have a solution. In this case, the electron can ionize early after the laser field is turned on, i.e., $t_{{\rm i},k} = 0$. At this time, the saddle point is located at $\tilde{r}_{{\rm s},k} (0)=(Z_k/\omega^2)^{1/3}$ and its height is $\tilde{h}_{{\rm s},k} (0) = -3 (Z_k \omega)^{2/3}/2$. For instance, this would be the case for the first ionization potential of $\mathrm{Li}$ subjected to a laser of wavelength $\lambda = 780 \; \mathrm{nm}$. 
\par
For $\mu_k \geq 1/4$, Eqs.~\eqref{eq:saddle_point_ionization_RF} have a solution. Equations~\eqref{eq:saddle_point_ionization_RF} can be rewritten as
\begin{subequations}
\begin{eqnarray}
\label{eq:effective_amplitude_general_potential_xstar}
&&  \mathcal{E}_k = \frac{\omega^2}{2}\tilde{r}_{{\rm s},k}^2 -\frac{2Z_k}{\tilde{r}_{{\rm s},k}} ,  \\
\label{eq:effective_amplitude_general_potential_Estar}
&& \tilde{E}^{\star}_k = -\omega^2 \tilde{r}_{{\rm s},k} +\frac{Z_k}{\tilde{r}_{{\rm s},k}^2} ,
\end{eqnarray}
\end{subequations}
where $\tilde{E}^{\star}_k=E_0 f(t_{{\rm i},k})$ is the effective amplitude of the laser field at ionization of the electron $\mathrm{e}_k^-$.
Equation~\eqref{eq:effective_amplitude_general_potential_xstar}, in addition to the constraints on the existence of a solution imposed by Eqs.~\eqref{eq:saddle_point_ionization_RF}, has a solution given by
\begin{equation}
\label{eq:position_saddle_point_Ip}
\tilde{r}_{{\rm s},k} =\left( \dfrac{Z_k}{\omega^2} \right)^{1/3} \min \left[ 1,\rho (\mu_k)\right] , 
\end{equation}
where $\rho (\mu_k) = 2^{1/3}\left[(\sqrt{\mu_k + 1} + 1 )^{1/3}- (\sqrt{\mu_k+1}-1)^{1/3}\right]$. The condition $\mu_k$ larger than $1/4$ corresponds to the condition $\rho (\mu_k)$ larger than one. Substituting Eq.~\eqref{eq:position_saddle_point_Ip} in Eq.~\eqref{eq:effective_amplitude_general_potential_Estar}, we obtain the effective amplitude at ionization of the $k$-th electron
\begin{equation}
\label{eq:effective_amplitude}
\tilde{E}^{\star}_k = \left( Z_k \omega^4 \right)^{1/3} \max \left[ 0 , \dfrac{1 - \rho(\mu_k)^3}{\rho(\mu_k)^2} \right] .
\end{equation}
In order to ionize over the barrier, the peak amplitude of the laser field $E_0$ must be greater than the effective amplitude of the laser at which the ionization of the electron can occur. As a result, the intensity at which the electron can ionize over the barrier is given by
$$
\tilde{I}_{\mathrm{OTB}, k} = 2 Z_k^{2/3} \omega^{8/3} \left[ \max \left( 0,   \dfrac{1 - \rho(\mu_k)^3}{\rho(\mu_k)^2} \right) \right]^2.
$$
For intensities $I < \tilde{I}_{\mathrm{OTB}, k}$, the $k$-th electron cannot ionize over the barrier, and remains topologically bounded by the RF-effective potential. For intensities $I > \tilde{I}_{\mathrm{OTB}, k}$, the $k$-th electron can potentially ionize over the barrier if it is dynamically allowed.
\par
We remark that this threshold intensity strongly depends on the laser frequency. Increasing the laser frequency so that $\mu_k < 1/4$ allows for potentially getting over-the-barrier ionization regardless of the intensity of the laser. We notice that for large values of $\mu_k$, i.e., for $\vert {\cal E}_k\vert ^3\gg (Z_k \omega)^2$, the value of $\tilde{E}_k^\star$ tends to the one obtained from the adiabatic approximation in the LF [see Eq.~\eqref{eq:effective_laser_amplitude_adiabatic}]. More precisely, we have
$$
\tilde{I}_{{\rm OTB},k}=I_{{\rm OTB},k}\left[1-8\frac{(Z_k\omega)^2}{\vert {\cal E}_k\vert^3} \right] + {\rm O}\left(\frac{(Z_k\omega)^4}{\vert {\cal E}_k\vert^6} \right) ,
$$
where $I_{{\rm OTB},k}$ is the intensity threshold for over-the-barrier ionization in the adibatic approximation in the LF [see Eq.~\eqref{eq:ImaxLF}].
For $\mathcal{E}_1 \approx - 0.28$ (close to $\mathrm{Mg}$) and $\lambda = 780 \; \mathrm{nm}$, the value of the nonadiabatic parameter is $\mu_1\approx 0.5$, and $\tilde{I}_{\mathrm{OTB},1} \approx 5.3 \times 10^{12} \; \mathrm{W}\cdot\mathrm{cm}^{-2}$ instead of ${I}_{\mathrm{OTB},1}=2.7 \times 10^{13} \; \mathrm{W} \cdot \mathrm{cm}^{-2}$ in the adiabatic approximation in the LF.
In the experimental measurements of Ref.~\citep{Gillen2001}, the enhancement in the DI probability is observed for intensities $I > \tilde{I}_{\mathrm{OTB},1}$, so the regime of intensities is compatible with over-the-barrier ionizations. For $\mathcal{E}_1 \approx - 0.6$ (close to $\mathrm{Ar}$) and $\lambda = 780 \; \mathrm{nm}$, the nonadiabatic parameter is $\mu_1\approx 4.7$ and the intensity is $\tilde{I}_{\mathrm{OTB},1} \approx 5.0 \times 10^{14} \; \mathrm{W}\cdot\mathrm{cm}^{-2}$, instead of ${I}_{\mathrm{OTB},1} \approx 5.7 \times 10^{14} \; \mathrm{W}\cdot\mathrm{cm}^{-2}$ in the adiabatic approximation in the LF. In the experimental measurements of Ref.~\citep{Pfeiffer2011}, the ionization times are measured for intensities $I > \tilde{I}_{\mathrm{OTB},1}$. As a consequence, we argue that the dominant ionization mechanism is also over-the-barrier in this set-up. 
Of course, close to the over-the-barrier ionization, there is a significant contribution of tunneling. In addition, since the barrier is thinner in the RF (in the adiabatic approximation), it increases the contribution of tunneling, since its probability exponentially increases with decreasing the width of the potential barrier. In any case, over-the-barrier ionization, when the channel is open, occurs with a high probability since no potential barrier can block the electron.

\begin{figure*}
	\includegraphics[width=.9\textwidth]{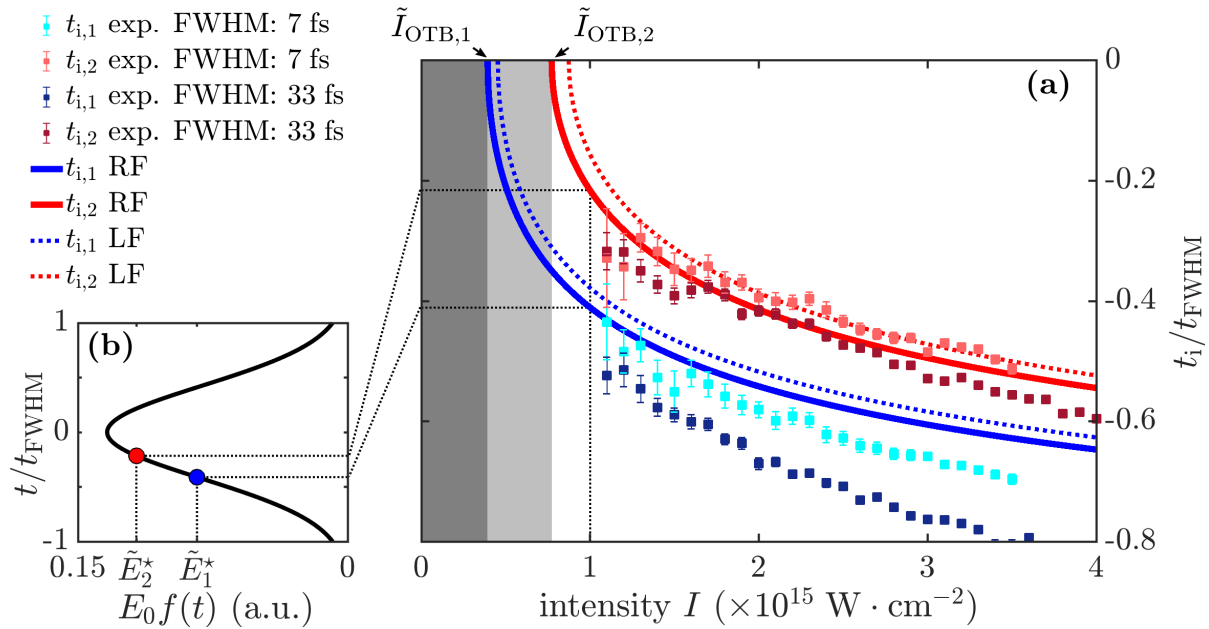}
	\caption{Ionization times of the first and second electron in the double ionization of $\mathrm{Ar}$ for a Gaussian laser envelope of the form $f(t) = \exp [- \ln (2) (2 t/t_{\rm FWHM})^2]$. The experimental data (squares) for the ionization time of the first and second electrons are reproduced from Ref.~\cite{Pfeiffer2011} for a laser ellipticity $\xi = 0.77$ (near-CP pulses). The laser amplitude is such that $E_0 = \sqrt{I / (\xi^2+1)}$. The solid and dotted lines are the theoretical classical predictions given by Eq.~\eqref{eq:ti_prediction} and Eq.~\eqref{eq:effective_laser_amplitude_adiabatic}, respectively, for a laser wavelength $\lambda = 788 \; \mathrm{nm}$, $\mathcal{E}_1 = - 0.6$, $Z_1 = 1$, $\mathcal{E}_2 = - 1$, $Z_2 = 2$~\citep{Wang2012}. The experimental measurements are given for $t_{\rm FWHM} = 33 \; \mathrm{fs}$ for a laser wavelength $\lambda = 788 \; \mathrm{nm}$ and for $t_{\rm FWHM} = 7 \; \mathrm{fs}$ for a laser wavelength $\lambda = 740 \; \mathrm{nm}$.}
	\label{fig:ionization_time_theory_experiments}
\end{figure*}

The time $t_{{\rm i},k}$ at which the electron $\mathrm{e}_k^-$ is ionized is controlled by the laser envelope and is obtained from the condition $\tilde{E}_k^{\star} = E_0 f ( t_{{\rm i},k} )$. We obtain
\begin{equation}
\label{eq:ti_prediction}
t_{{\rm i},k} = f^{-1} \left( \dfrac{\tilde{E}^{\star}_k}{E_0} \right) ,
\end{equation}
where $f^{-1}$ is the inverse function of the laser envelope. Since in general the function $f$ is non-injective, the ionization time $t_{{\rm i},k}$ corresponds to the smallest time given by Eq.~\eqref{eq:ti_prediction} [such that the effective amplitude $E_0 f(t)$ is increasing]. 
\par 
In the large $\mu_k$ regime, the value of the ionization times given by Eq.~\eqref{eq:ti_prediction} gives the same value as in Eq.~\eqref{eq:effective_laser_amplitude_adiabatic}. Its expansion is given by
\begin{equation}
\label{eq:tiNA}
f ( t_{{\rm i},k} ) = \frac{E_k^{\star}}{E_0} \left[ 1 - 4 \frac{( Z_k \omega )^2}{\vert {\cal E}_k\vert^3} \right] + O\left(\frac{( Z_k \omega )^4}{\vert {\cal E}_k\vert^6}\right) ,
\end{equation}
where $E_k^{\star}$ is the effective amplitude of the laser field in the adiabatic approximation in the LF [see Eq.~\eqref{eq:effective_laser_amplitude_adiabatic}]. The second term on the right-hand side of the equation quantifies the nonadiabatic effects on the ionization times. 
\par 
The existence of invariant tori (see Fig.~\ref{fig:poincare_section}a for instance) induces delays in the ionization of the electrons. As a consequence, the ionization times given by Eq.~\eqref{eq:ti_prediction} correspond to a lower bound of the times at which the electron ionizes. 
Figure~\ref{fig:ionization_time_theory_experiments} shows the ionization times of the electrons in $\mathrm{Ar}$ ($\mathcal{E}_1 = -0.6$ and $\mathcal{E}_2 = -1$~\citep{Wang2012}) as a function of the laser intensity from experiments~\citep{Pfeiffer2011} and the classical prediction given by Eq.~\eqref{eq:ti_prediction} for a Gaussian laser pulse with full width at half maximum (FWHM) of $7 \; \mathrm{fs}$ and $33 \; \mathrm{fs}$. Our prediction given by Eq.~\eqref{eq:ti_prediction} agrees qualitatively well with the ionization times measured experimentally. The quantitative discrepancy comes from tunnel ionization which occurs before over-the-barrier ionization. In sum, the comparison of Eq.~\eqref{eq:ti_prediction} with the experimental measurements of Ref.~\citep{Pfeiffer2011} indicates that the ionization process is well described by looking at the times when the electron ionizes over the barrier. For these laser parameters, the subcycle nonadiabatic effects are rather small, and as a consequence, the adiabatic approximations in the LF and in the RF provide similar predictions. We expect nonadiabatic effects to be more evident if the laser wavelength is decreased, e.g., by a factor 2. The nonadiabatic effects include a decreasing of the ionization times with decreasing wavelength. 

\section{Recollisions for nonsequential double ionization \label{sec:NSDI_mechanisms}}

\begin{figure*}
	\centering
	\includegraphics[width=\textwidth]{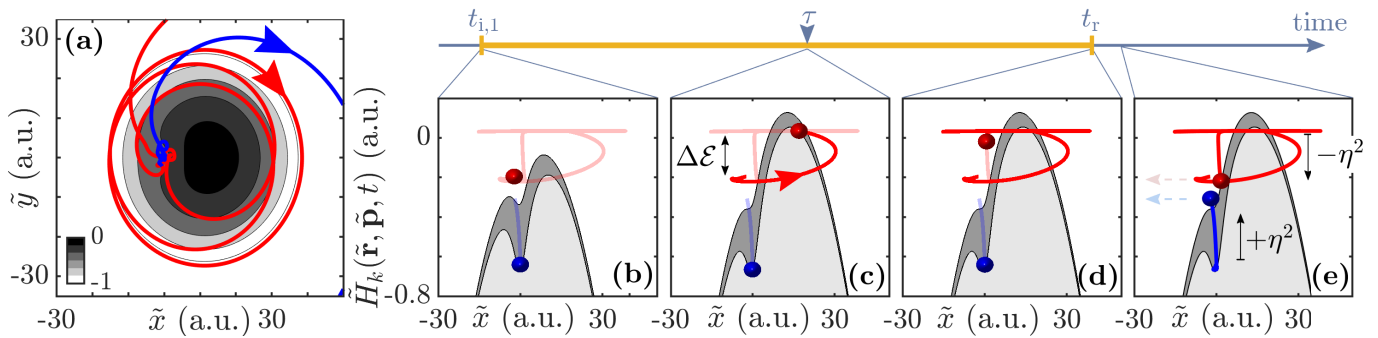}
	\caption{Typical NSDI of Hamiltonian~\eqref{eq:Hamiltonian_NSDI_LF} induced by a recollision in the RF for Mg ($\mathcal{E}_g=-0.83$, $a=3$), $I = 10^{14} \unites{W}\cdot \mathrm{cm}^{-2}$ and a trapezoidal envelope 2--4--2 (duration of ramp-up $\tau = 2 T$). The red and blue curves are the recolliding and bound electron trajectories for $t \in [t_{{\rm i},1} , 5.7 T]$, respectively. (a) Projection in the plane $(\tilde{x},\tilde{y})$. A contour plot of the RF-effective potential $V_{{\rm RF},1} (\tilde{\mathbf{r}},\tilde{\mathbf{p}} ,\tau)$ is displayed on the leftmost panel. (b--e) Snapshots of the electron in the plane $( \tilde{x} ,\tilde{H}_k (\tilde{\mathbf{r}},\tilde{\mathbf{p}},t))$. The dark and light gray areas are the classical forbidden regions delimited by the RF-effective potential of the recolliding electron $V_{{\rm RF},1} (\tilde{x} \hat{\tilde{\mathbf{x}}},t)$ and of the bound electron $V_{{\rm RF},2} (\tilde{x} \hat{\tilde{\mathbf{x}}},t)$, respectively. The red and blue dots are the recolliding and bound electrons at times (b) $t=t_{{\rm i},1} \approx 0.9 T$, (c) $t=\tau = 2T$, (d) $t=t_{\rm r} \approx 4.5 T$ and (e) $t \approx 4.6 T$. (d) The light red and blue dashed arrows indicate the subsequent ionization of the electrons.}
	\label{fig:scenario_NSDI}
\end{figure*}

The second class of double ionizations is the non-sequential double ionization. This double ionization channel, and its associated recollision mechanism, is extensively documented for linear polarization in the literature. Much less is known for circular or near-circular polarizations. \par
One typical NSDI trajectory of Hamiltonian~\eqref{eq:Hamiltonian_NSDI_LF} is depicted in Fig.~\ref{fig:scenario_NSDI} in the RF. The red electron leaves the core at time $t_{{\rm i},1} < \tau$ early after the laser field is turned on, gains energy during the ramp-up, and then returns to the core at time $t_{\rm r}$. Close to the core, the red and blue electrons exchange energy, and both ionize. In this section, we identify the conditions under which recollisions can trigger NSDI. In particular, we determine the intensity which allows the electron to bring back a sufficient amount of energy to the core. Then, we identify the laser wavelengths and the atoms at which manifestations of these recollisions can be observed.

\subsection{Minimum intensity for NSDI\label{sec:minimum_intensity}}

\begin{figure}
	\centering
	\includegraphics[width=.5\textwidth]{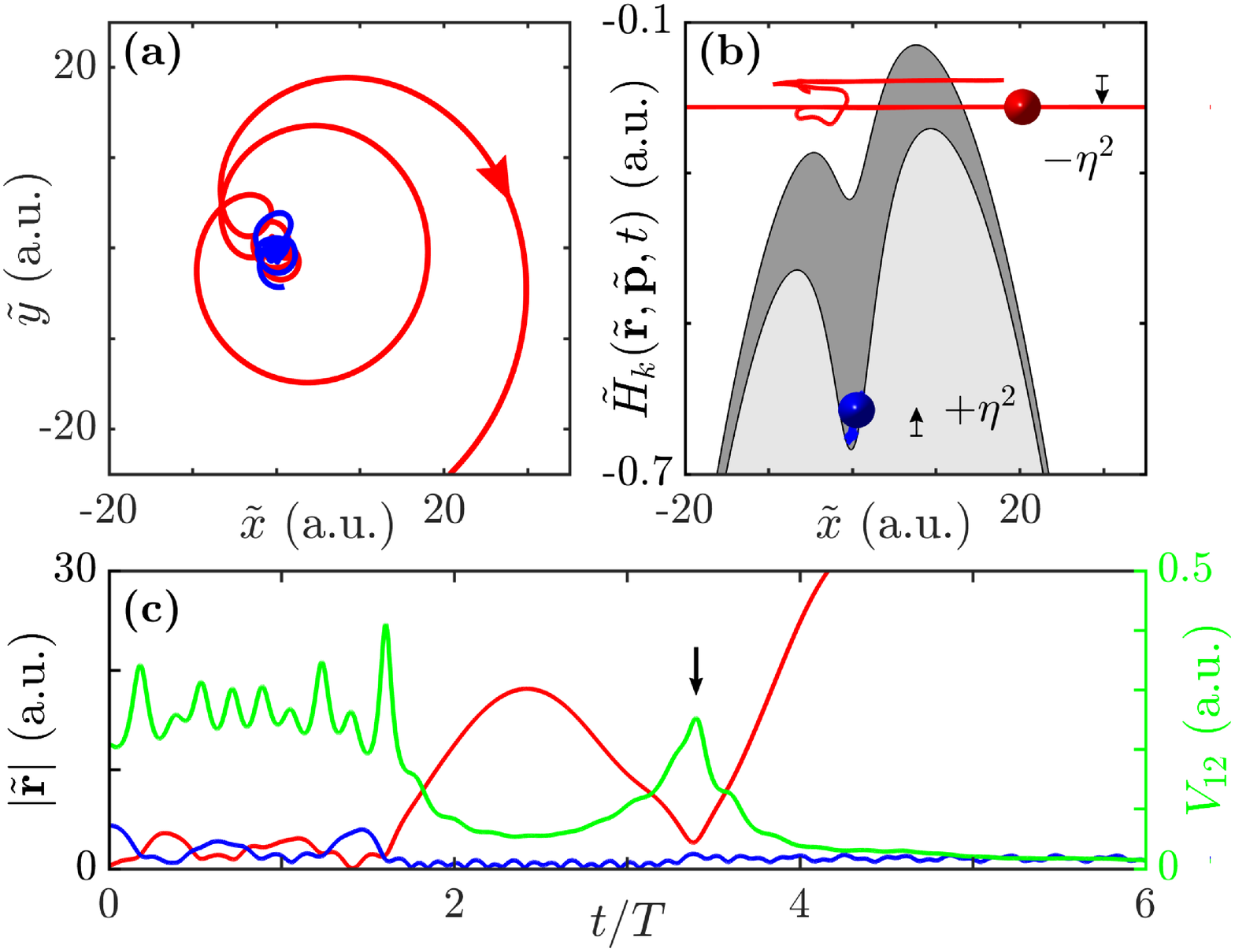}
	\caption{Typical recollision without NSDI of Hamiltonian~\eqref{eq:Hamiltonian_NSDI_LF} for $\mathrm{Mg}$ ($\mathcal{E}_g = -0.83$, $a=3$), $I = 10^{13} \; \mathrm{W} \cdot \mathrm{cm}^{-2}$ and a trapezoidal envelope 2--4--2 ($\tau=2T$). (a) Projection onto the polarization plane in the RF $(\hat{\tilde{\mathbf{x}}} , \hat{\tilde{\mathbf{y}}})$. (b) Projection onto the plane $(\tilde{x}, \tilde{H}_k (\tilde{\mathbf{r}} , \tilde{\mathbf{p}} , t))$. The gray surfaces are the RF-effective potential of the red (dark gray) and blue (light gray) electrons. (c) Distance as a function of time per laser cycle, and interaction between electrons $V_{12} = (|\tilde{\mathbf{r}}_1 - \tilde{\mathbf{r}}_2|^2+1)^{-1/2}$ as a function of time per laser cycle. The arrow on the lower panel indicates a recollision.}
	\label{fig:recollision_without_NSDI}
\end{figure}

After ionization and before it returns to the core, the recolliding electron must gain energy from the CP pulse to trigger NSDI. For instance, Fig.~\ref{fig:recollision_without_NSDI} shows typical trajectories of Hamiltonian~\eqref{eq:Hamiltonian_NSDI_LF} in the RF, where one electron recollides without triggering NSDI. The ionization occurs at time $t_{{\rm i},1} \approx 2T$ (so rather late, compared to Fig.~\ref{fig:scenario_NSDI}), which corresponds to the end of the ramp-up. When the electron returns to the core, at time $t_{\rm r} \approx 5T$ (indicated by an arrow in Fig.~\ref{fig:recollision_without_NSDI}), the electrons exchange an amount of energy denoted $\eta^2$. The energy gained by the recolliding electron is not sufficient to allow an effective transfer of energy to the inner electron so that its overcomes its RF-effective potential. Consequently, only one electron ionizes. The energy gained by the recolliding electron from the CP pulse during its excursion outside the core region must be sufficiently large to allow both electrons to overcome their respective RF-effective potential, i.e., to get an energy larger than the energy of the saddle points of the RF-effective potentials.

First, we consider the case where the electron ${\rm e}_1^-$ ionizes and returns during the plateau, i.e., when $f(t)=1$. The recollision conditions given by $\tilde{\mathbf{r}} (t_{{\rm i},1}) \approx \tilde{\mathbf{r}} (t_{\rm r}) \approx \boldsymbol{0}$ and the conservation of energy implies that $\tilde{H}_1 (\boldsymbol{0} , \tilde{\mathbf{p}} (t_{{\rm i},1}),t_{{\rm i},1}) \approx \tilde{H}_1 (\boldsymbol{0} , \tilde{\mathbf{p}} (t_{\rm r}),t_{{\rm r}})$. Therefore, the energy difference of the recolliding electron $\Delta \mathcal{E} = |\tilde{\mathbf{p}} (t_{\rm r})|^2/2 - |\tilde{\mathbf{p}} (t_{{\rm i},1})|^2 / 2$ is $\Delta \mathcal{E} \approx 0$.
In contrast to recolliding electrons driven by LP pulses, recolliding electrons driven by CP pulses cannot gain energy from the laser if the envelope of the field does not vary between ionization and return, as exemplified by Fig.~\ref{fig:recollision_without_NSDI} compared to Fig.~\ref{fig:scenario_NSDI}.
\par
As a consequence, the laser envelope has to be taken into account. In order to determine the conditions under which the recolliding electron gains energy from the CP pulse, we neglect the Coulomb potential, and we use the slowly varying envelope approximation. At first order, the energy difference of the recolliding electron is
\begin{equation}
\label{eq:energy_gain_recolliding_electron_slowly_varying_envelope}
\Delta \mathcal{E} \approx \frac{E_0^2}{2\omega^2} \left[ f (t_{\rm r})^2 - f(t_{{\rm i},1})^2 \right].
\end{equation}
The only way for the electron to gain energy is to return to the core when the laser envelope is greater than it was at ionization. Therefore, the ionization time must take place during the ramp-up so that the energy of the recolliding electron can be boosted by the variations of the laser envelope, as observed in Fig.~\ref{fig:scenario_NSDI}, and return to the core before its effective intensity has decreased, e.g., during the plateau.
\par
Right before it returns to the core, at time $t_{\rm r}^-$, the energy of the recolliding (outer) electron is $\tilde{H}_1 ( t_{\rm r}^- ) \approx \tilde{H}_1 ( t_{{\rm i},1}) + \Delta \mathcal{E}$. Here $\tilde{H}_1 ( t_{\rm r}^- )$ denotes the value of the Hamiltonian at $t=t_{\rm r}^-$, i.e., $\tilde{H}_1 ( t_{\rm r}^- )=\tilde{H}_1 (\tilde{\mathbf{r}}(t_{\rm r}^-) ,\tilde{\mathbf{p}}(t_{\rm r}^-), t_{\rm r}^-)$ for simplicity. The energy of the inner electron does not vary as much as the one of the recolliding electron because the time-dependent term in Hamiltonian~\eqref{eq:Hamiltonian_RF_1ae} is small compared to the energy of the inner electron, i.e., $\tilde{H}_2 ( t_{\rm r}^-) \approx \tilde{H}_2 (t_{{\rm i},1})$. Initially, the atom is in the ground state of energy $\mathcal{E}_g$, and using Eq.~\eqref{eq:Hamiltonian_NSDI_screen_effect}, one gets $\tilde{H}_1 ( t_{{\rm i},1}) + \tilde{H}_2 ( t_{{\rm i},1}) \approx \mathcal{E}_g$. The amount of energy exchanged by the two electrons (via the electron-electron interaction term) is denoted $\eta^2$. After the exchange of energy, at time $t_{\rm r}^+$, the energy of the outer electron becomes $\tilde{H}_1 ( t_{\rm r}^+) \approx \tilde{H}_1 (t_{\rm r}^-) - \eta^2$ and the energy of the inner electron becomes $\tilde{H}_2 (t_{\rm r}^+) \approx \tilde{H}_2 ( t_{\rm r}^-) + \eta^2$. We consider the case where the recolliding electron has high velocity compared to the inner electron, therefore at $t_r^+$, ${\rm e}_1^-$ is far from the core, so its effective charge is screened by $1$. In order to both ionize, the energy of both electrons must be greater than the energy of their saddle point~\citep{Kamor2013}, i.e., $\tilde{H}_k ( t_{\rm r}^+) > \tilde{h}_{{\rm s},k}( t_{\rm r})$. Consequently, a necessary condition for which both electrons ionize is
\begin{equation}
\label{eq:sufficient_energy_gined_to_trigger_NSDI_condition}
\Delta \mathcal{E} \gtrsim \tilde{h}_{{\rm s},1}( t_{\rm r}) + h_{{\rm s},2}( t_{\rm r}) - \mathcal{E}_g .
\end{equation}
Substituting Eq.~\eqref{eq:energy_gain_recolliding_electron_slowly_varying_envelope} with $f(t_{\rm r}) = 1$ and $\tilde{E}_1^{\star} = E_0 f(t_{{\rm i},1})$ in Eq.~\eqref{eq:sufficient_energy_gined_to_trigger_NSDI_condition}, we obtain the minimum intensity for which we expect that NSDI can be triggered by recollisions
\begin{equation}
\label{eq:minimum_intensity_definition}
\tilde{I}_{\min} \approx \tilde{I}_{\rm OTB,1} + 4 \omega^2 \left[ \tilde{h}_{{\rm s},1}( t_{\rm r}) + \tilde{h}_{{\rm s},2}( t_{\rm r}) - \mathcal{E}_g \right] .
\end{equation}
Notice that the right-hand side of Eq.~\eqref{eq:minimum_intensity_definition} also depends on the intensity in the terms $\tilde{h}_{{\rm s},1}( t_{\rm r})$ and $\tilde{h}_{{\rm s},2}( t_{\rm r})$. The minimum intensity given by Eq.~\eqref{eq:minimum_intensity_definition} must be computed numerically using Eq.~\eqref{eq:fixed_point_definition} and Eq.~\eqref{eq:hRFk} with $f(t_r)=1$. 

\begin{figure*}
	\centering
	\includegraphics[width=\textwidth]{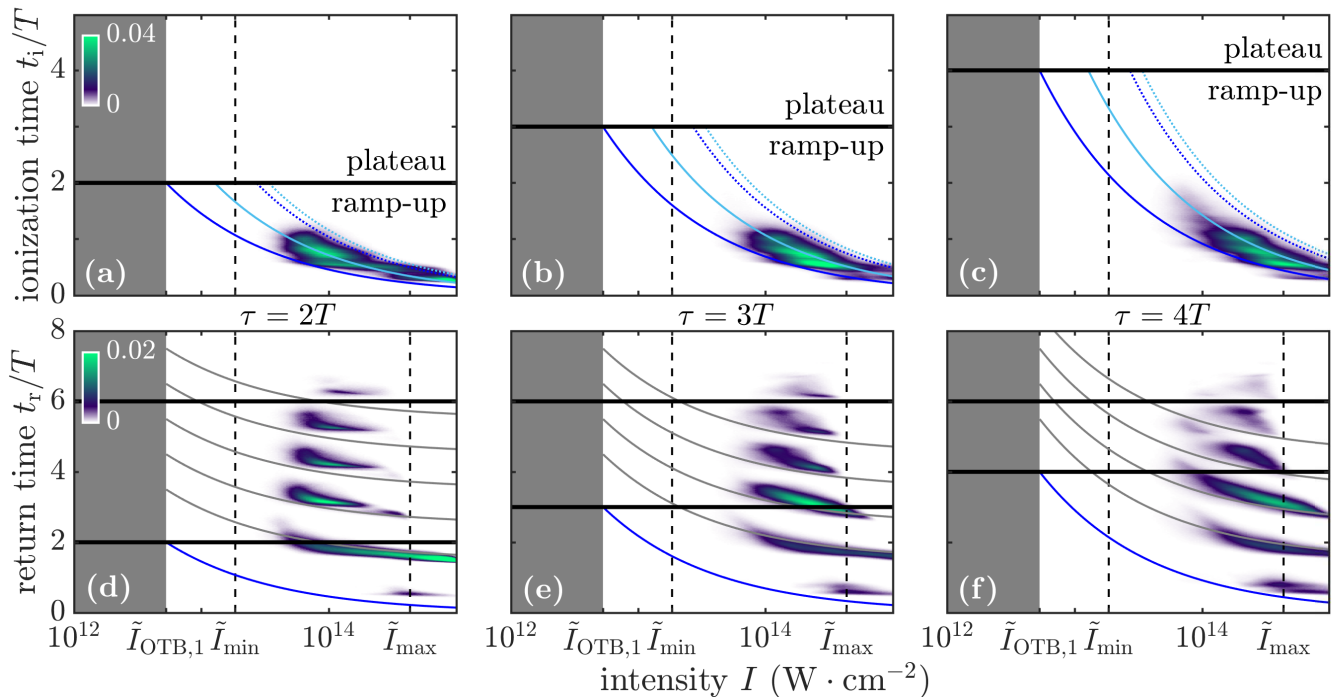}
	\caption{Probability density function of the ionization time $t_{{\rm i},1}$ [panels (a--c)] and return time $t_{{\rm r},1}$ [panels (d--f)] of the recolliding electrons triggering NSDI of Hamiltonian~\eqref{eq:Hamiltonian_NSDI_LF} as a function of the laser intensity $I$ for Mg ($\mathcal{E}_g = -0.83$, $a=3$). The parameters are the same as in Figs.~\ref{fig:probability_curves}. The ramp-up duration is $\tau$, the plateau duration is $6T-\tau$ and the ramp-down is of duration $2T$ (indicated by horizontal black solid lines): in (a) and (d) $\tau = 2T$, in (b) and (e) $\tau = 3T$, in (e) and (f) $\tau = 4T$. The blue dashed and solid curves are the theoretical prediction of the ionization time of the first electron for the hard-Coulomb potential given by Eq.~\eqref{eq:effective_laser_amplitude_adiabatic} (adiabatic approximation in the LF) and Eq.~\eqref{eq:ti_prediction} (adiabatic approximation in the RF), respectively. The light blue dashed and solid curves are the theoretical prediction of the ionization time of the first electron for the soft-Coulomb potential with $a=3$. The vertical dashed black curves are $I = \tilde{I}_{\min}$ [see Eq.~\eqref{eq:minimum_intensity_definition}]. The gray regions are intensities $I < \tilde{I}_{\rm OTB,1}$. The vertical dashed black line indicates $I = \tilde{I}_{\max}$ given by Eq.~\eqref{eq:exact_Imax_definition}. The gray solid curves are given by $t_{\rm r} = t_{{\rm i},1} + (n + 1/2) T$, with $n = 1, 2, 3,4,5$.}
	\label{fig:ionization_time}
\end{figure*} 

In Fig.~\ref{fig:ionization_time}a--c, the leftmost vertical black dashed curve is the prediction of the minimum laser intensity $\tilde{I}_{\min}$ necessary for triggering NSDI as given by Eq.~\eqref{eq:minimum_intensity_definition}. The color scale is the distributions of ionization times $t_{{\rm i},1}$ of the electron $\mathrm{e}_1^-$ which undergoes a recollision and triggers NSDI of Hamiltonian~\eqref{eq:Hamiltonian_NSDI_LF}. We observe that the prediction for $\tilde{I}_{\rm min}$ given by Eq.~\eqref{eq:minimum_intensity_definition} agrees well with the minimum intensity at which we observe recollisions triggering NSDI. In addition, as we observe in Fig.~\ref{fig:ionization_time}a--c, $\tilde{I}_{\min} > \tilde{I}_{\rm OTB,1}$. The electron needs to ionize early during the ramp-up in order to bring back a sufficient amount of energy to trigger NSDI, and as a consequence, undergoes an envelope-driven recollision
~\cite{Dubois2020}.

\subsection{Envelope-driven recollisions: ionization and return times}

Envelope-driven recollisions correspond to electrons which gain energy through the variations of the laser envelope~\cite{Dubois2020} between ionization and return to the core. In this section, we detail the various steps of an envelope-driven recollision which triggers NSDI in CP pulses. In Fig.~\ref{fig:scenario_NSDI}, we plotted a typical recollision trajectory of Hamiltonian~\eqref{eq:Hamiltonian_NSDI_LF} in the RF between ionization time $t_{{\rm i},1}$ of the first electron and its return time $t_{\rm r}$.
\par
For $t < t_{{\rm i},1}$ (see Fig.~\ref{fig:scenario_NSDI}b), the time-dependent term in Hamiltonian~\eqref{eq:Hamiltonian_NSDI_LF} is small compared to the ground state energy of the electrons, therefore the energy of the system is almost conserved, i.e., $H ( \mathbf{r}_1 , \mathbf{p}_1 , \mathbf{r}_2, \mathbf{p}_2 , t ) \approx \mathcal{E}_g$.
In the RF, the same approximation holds. We assume that the energy of the two electrons is such that $\tilde{H}_k (\tilde{\mathbf{r}},\tilde{\mathbf{p}},t_{{\rm i},1}) \approx \mathcal{E}_k$. At $t=t_{{\rm i},1}$, where $t_{{\rm i},1}$ is given by Eq.~\eqref{eq:ti_prediction}, the height of the potential barrier is low enough for the electron $\mathrm{e}_1^-$ to ionize. In Fig.~\ref{fig:ionization_time}a--c, the blue solid and dashed lines are the prediction of the ionization time given by Eq.~\eqref{eq:ti_prediction} (adiabatic approximation in the RF) and Eq.~\eqref{eq:effective_laser_amplitude_adiabatic} (adiabatic approximation in the LF), respectively. The blue and light blue curves are for the hard-Coulomb potential and soft-Coulomb potential with $a=3$. We observe a good qualitative agreement between the solid curves and the distribution of the ionization times computed numerically for the two-electron Hamiltonian~\eqref{eq:Hamiltonian_NSDI_LF}. In addition, we observe a nonadiabatic effect by comparing the dashed and the solid curves: Ionization happens earlier by taking into account the energy variations of the electron on subcycle timescales happening in the LF. This is consistent with Eq.~\eqref{eq:tiNA}.
We recall that the ionization times are predicted by neglecting the dynamics. As a consequence, the ionization times obtained from Eqs.~\eqref{eq:ti_prediction}-\eqref{eq:effective_laser_amplitude_adiabatic} correspond to a lower bound of the ionization times of the first electron, since its escape might be obstructed by dynamical structures like invariant tori. From Fig.~\ref{fig:ionization_time}a--c, we conclude that it is reasonable to assume that, at a given intensity, the first electrons all ionize at approximately the same time, and rather early in the ramp-up of the pulse, in order to trigger NSDI. 

\begin{figure}
	\centering
	\includegraphics[width=0.5\textwidth]{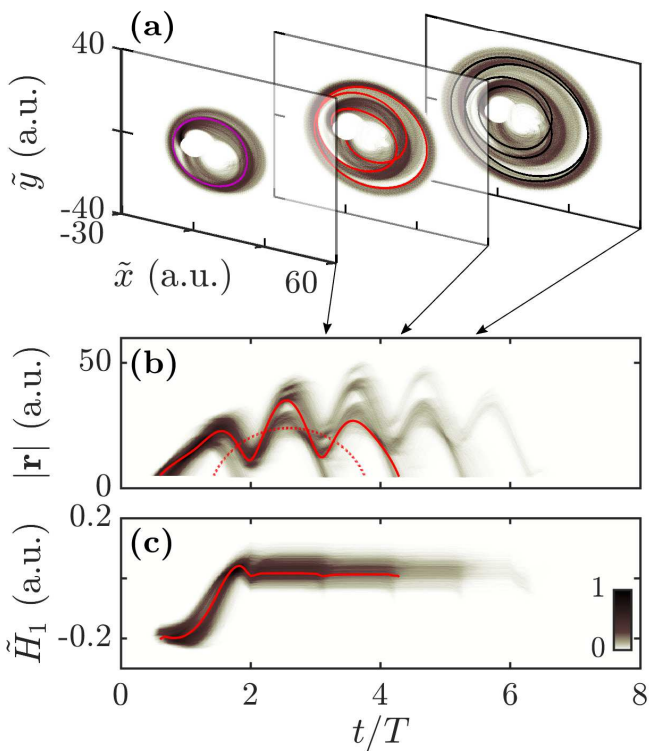}
	\caption{Distributions of the recolliding trajectories leading to NSDI of Hamiltonian~\eqref{eq:Hamiltonian_NSDI_LF} between their ionization time $t_{{\rm i}}$ and their return $t_{{\rm} r}$ (where $|\mathbf{r} (t_{{\rm i}})| = |\mathbf{r} (t_{{\rm r}})| = 5$) for Mg ($\mathcal{E}_g = -0.83$, $a=3$), with a 2--4--2 trapezoidal envelope ($\tau=2T$) and $I=10^{14} \; \mathrm{W}\cdot \mathrm{cm}^{-2}$. The parameters are the same as in Fig.~\ref{fig:scenario_NSDI}. (a) Distribution of the trajectories in the plane $(\tilde{x},\tilde{y})$ for return times (from left to right) $t_{\rm r} \in [2.5T,3.75T]$, $t_{\rm r} \in [3.75T,5T]$ and $t_{\rm r} \in [5T,6T]$. The purple and black curves are RPOs of Hamiltonian~\eqref{eq:Hamiltonian_RF_1ae}. Distribution of (a) the distance $|\mathbf{r}|$ and (b) Jacobi value $\tilde{H}_1$ of the recolliding electron as a function of time. (a)--(c) The red curve is a typical recolliding trajectory leading to NSDI of Hamilonian~\eqref{eq:Hamiltonian_NSDI_LF}. The dotted trajectory in (b) corresponds to the guiding-center trajectory of the recolliding electron whose dynamics is governed by Hamiltonian~\eqref{eq:Hamiltonian_isochrone}. }
	\label{fig:distributions_trajectories}
\end{figure}

\begin{figure}
	\centering
	\includegraphics[width=0.5\textwidth]{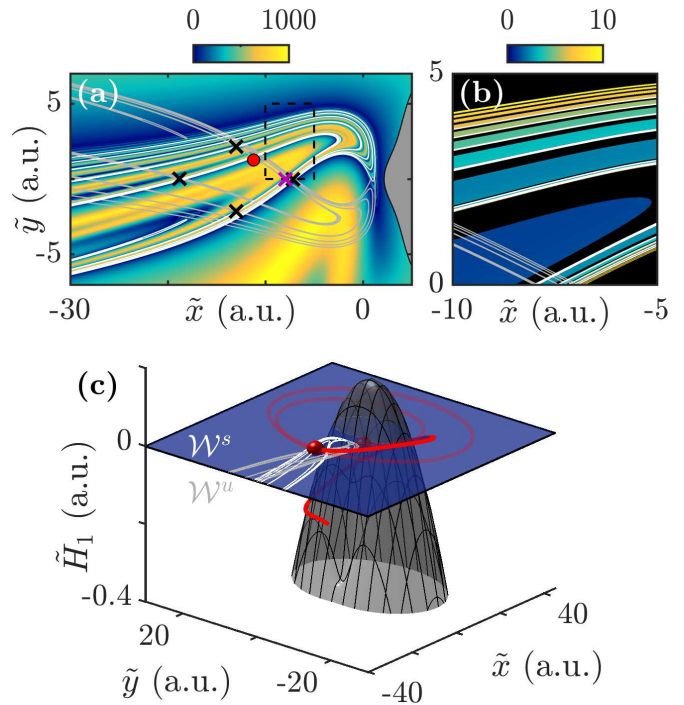}
	\caption{Stable $\mathcal{W}^s$ (white solid lines) and unstable $\mathcal{W}^u$ (gray solid lines) manifolds of fixed point [purple cross in panel (a)] under the Poincar\'{e} section $\dot{\tilde{x}}=0$ and $\ddot{\tilde{x}} > 0$ of Hamiltonian~\eqref{eq:Hamiltonian_RF_1ae} in the plane $(\tilde{x},\tilde{y})$ for $I = 10^{14} \; \mathrm{W}\cdot \mathrm{cm}^{-2}$, $f=1$, $a=3$, $\tilde{H}_k (\tilde{\mathbf{r}},\tilde{\mathbf{p}}) = 0$, $k=1$.  $(\tilde{x},\tilde{y})$. 
	(a) Final distance of the electron as a function of the initial conditions $(\tilde{x},\tilde{y})$ on the Poincar\'e section. The purple and black crosses are fixed points under this Poincar\'{e} section corresponding to the purple and black periodic orbits depicted in Fig.~\ref{fig:distributions_trajectories}a. The gray region is the forbidden region on this Poincar\'{e} section. The trajectories are integrated for ten laser periods with a constant laser envelope $f=1$. 
	(b) Time when the electron crosses $|\mathbf{r}| = 5$ as a function of the initial conditions $(\tilde{x},\tilde{y})$ in the dashed square of panel (a). 
	(c) The gray surface is the RF-effective potential at the end of the ramp-up $V_{{\rm RF},1} (\tilde{\mathbf{r}},\tau)$. The red solid line is a a typical recolliding trajectory leading to NSDI of Hamiltonian~\eqref{eq:Hamiltonian_NSDI_LF} in the RF (same trajectory as the red one depicted in Figs.~\ref{fig:distributions_trajectories}) for Mg ($a=3$ and $\mathcal{E}_g=-0.83$), $I=10^{14} \; \mathrm{W}\cdot \mathrm{cm}^{-2}$ and  $\tau=2T$. The solid red curve is the trajectory before the end of the ramp-up. The location of the electron at the end of the ramp-up is indicated by a red ball [also indicated in panel (a) by a red dot]. The transparent red curve is the trajectory after the end of the ramp-up. The transparent red ball is the location of the electron at the return time. }
	\label{fig:invariant_manifolds}
\end{figure}

At $t_{{\rm i},1} < t < t_{\rm r}$ (see Figs.~\ref{fig:scenario_NSDI}c--d), $\mathrm{e}_1^-$ is outside the ionic core region. In the RF, the electron spins around the RF-effective potential far from the core (see Fig.~\ref{fig:scenario_NSDI}a). During its excursion outside the core and before the end of the ramp-up (at time $t_{{\rm i},1} < t < \tau$), the time-dependent term in the reduced Hamiltonian~\eqref{eq:Hamiltonian_RF_1ae} becomes significant compared to its initial energy, and $\tilde{H}_1 ( \tilde{\mathbf{r}}, \tilde{\mathbf{p}} , t)$ starts varying significantly. During the ramp-up, the recolliding electron gains a significant amount of energy which comes from the variations of the laser envelope according to Eq.~\eqref{eq:energy_gain_recolliding_electron_slowly_varying_envelope}. This is observed in Fig.~\ref{fig:distributions_trajectories}c, which shows the distribution of the energy of $\mathrm{e}_1^-$ during the recollisions of Hamiltonian~\eqref{eq:Hamiltonian_NSDI_LF} leading to NSDI. We propose two complementary explanations for understanding how the electron returns to the parent ion: (i) After ionization, the electron trajectory is driven back to the ionic core by its guiding center trajectory with a bound motion~\cite{Dubois2018, Dubois2018_PRE, Dubois2020}, as shown in Fig.~\ref{fig:distributions_trajectories}b. (ii) After the end of the ramp-up, the electron populates regions in phase space where invariant structures drive the electron back to the core~\cite{Kamor2013}, as shown in Fig.~\ref{fig:invariant_manifolds}. In the meantime, $\mathrm{e}_2^-$ is trapped close to the core, inside the well of the effective potential $V_{{\rm RF},2}$, on invariant tori near the origin as observed in Fig.~\ref{fig:poincare_section}. These two explanations will be detailed in what follows. 
\par
Right after the return time $t_{\rm r}$ (see Figs.~\ref{fig:scenario_NSDI}d--e), the two electrons exchange energy. The energy of $\mathrm{e}_1^-$ decreases and the energy of $\mathrm{e}_2^-$ increases sufficiently to become larger than the energy of its saddle point. Then, both ionize. Figure~\ref{fig:ionization_time}d--f shows the distributions of return times $t_{\rm r}$ of $\mathrm{e}_1^-$ for trajectories of Hamiltonian~\eqref{eq:Hamiltonian_NSDI_LF} leading to NSDI. We observe that the recolliding electrons return to the core at some prescribed times, equally spaced by one laser period. This can be easily understood by using the mechanism (i): In the shifted coordinates $\tilde{\mathbf{r}}_{g} = \tilde{\mathbf{r}} - E_0 f(t) / \omega^2 \hat{\mathbf{x}}$ and $\tilde{\mathbf{p}}_{g} = \tilde{\mathbf{p}} - E_0 f(t) / \omega \hat{\mathbf{y}}$, 
also referred to as guiding-center coordinates (see Refs.~\cite{Mauger2010_PRL,Dubois2018_PRE} for more details), the dynamics of the guiding centers in the RF is given by~\cite{Dubois2018_PRE}
$$
H_{g,1} (\tilde{{\bf r}}_{g},\tilde{{\bf p}}_{g}) = \dfrac{|\tilde{{\bf p}}_{g}|^2}{2} + V_1 (\tilde{{\bf r}}_{g}) - \omega \tilde{{\bf r}}_{g} \times \tilde{{\bf p}}_{g} \cdot \hat{\mathbf{z}} .    
$$
In polar coordinates, the dynamics of the guiding center is determined by the distance from the origin $r_{g} = |\tilde{\mathbf{r}}_{g}|$, the angle $\theta_{g} = \tan^{-1} (\tilde{y}_{g}/\tilde{x}_{g})$, the radial momentum $p_{g,r} = \tilde{\mathbf{r}}_{g} \cdot \tilde{\mathbf{p}}_{g} /|\tilde{\mathbf{r}}_{g}|$ and the angular momentum $p_{g,\theta} = - \tilde{\mathbf{r}}_{g} \times \tilde{\mathbf{p}}_{g} \cdot \hat{\tilde{\mathbf{z}}}$. Since, for symmetric potentials, the angular momentum of the guiding center is conserved, we evaluate this quantity at ionization, i.e., $p_{\theta,g} \sim E_0^2 f(t_{{\rm i},k})^2/\omega^3$. For large $r_{g}$, of the order of the quiver radius and above, one has $p_{\theta , g} / r_g^2 \lesssim \omega f(t_{{\rm i},k})^2$. Since the electron ionizes rather early after the laser field is turned on, $p_{\theta,g}/r_{g}^2$ is rather small compared to the laser frequency $\omega$. The approximate Hamiltonian after ionization becomes
\begin{equation}
\label{eq:Hamiltonian_isochrone}
H_{g}\approx \frac{p_{g,r}^2}{2} - \frac{1}{r_{g}}+\omega p_{g,\theta} .
\end{equation}
The guiding-center trajectory is bounded if and only if $G_1 = p_{g,r}^2/2 - 1/r_{g}$ is such that $G_1 < 0$ (see Sec.~\ref{sec:critical_ionization_potential} for more details). The associated dynamics is an elliptical motion in the variables $(r_{g},p_{g,r})$, coupled to a rotation $\theta_g(t)=\omega t+\theta_0$ in the variables $(\theta_g,p_{\theta,g})$. The distance of the electron to the core is then
$$
r(t)= \left[ r_g(t)^2+\frac{E_0^2}{\omega^4}+2r_g(t)\frac{E_0}{\omega^2}\cos (\omega t +\theta_0) \right]^{1/2} ,
$$
which is an oscillation of period $T=2\pi/\omega$ around a Kepler orbit $r_g(t)$. This Kepler orbit brings the recolliding electron back to the core (otherwise the electron ionizes). Given the large oscillations around this Kepler orbit, the times of returns are at the minimum of the oscillations. Given that all electrons ionize approximately at the same time (for a given intensity), the return times are $t_r\approx t_0+nT$, where $n\in \mathbb{N}^*$, where $t_0$ is an additional delay depending on the pulse shape and the ionization time. This conclusion is in good agreement with the return times obtained from Hamiltonian~\eqref{eq:Hamiltonian_NSDI_LF} in Fig.~\ref{fig:ionization_time}.  
\par
A more accurate explanation is afforded by Hamiltonian nonlinear dynamics: The electron at the end of the ramp-up, populates regions in phase space where invariant structures drives it back to the core~\cite{Kamor2013}, as depicted in Fig.~\ref{fig:invariant_manifolds}c. These invariant structures are the stable and unstable manifolds of key periodic orbits, called recolliding periodic orbits~\cite{Kamor2013} (RPOs), such as the one depicted in purple in Fig.~\ref{fig:distributions_trajectories}a. They are located at an energy $\tilde{H}_1 (\tilde{\mathbf{r}},\tilde{\mathbf{p}})$ near the energy of the top of the RF-effective potential, i.e., around $E_0^2/(2\omega^2)$. In Fig.~\ref{fig:invariant_manifolds}a, we show the final distance of the electrons initiated on the Poincar\'{e} section $\dot{x} = 0$ and $\ddot{x} > 0$. The purple and black crosses are fixed points under this Poincar\'{e} section corresponding to the purple and black RPOs in Fig.~\ref{fig:distributions_trajectories}a, respectively. In Figs.~\ref{fig:invariant_manifolds}, the white and gray curves correspond to the stable $\mathcal{W}^s$ and unstable $\mathcal{W}^u$ manifolds of the purple cross on the Poincar\'e section. In Fig.~\ref{fig:invariant_manifolds}a, we observe regions of sensitivity with respect to the initial conditions (with its characteristic feature of stretching and folding), where recollisions take place. These regions are organized by $\mathcal{W}^s$. Near the intersections of $\mathcal{W}^s$ and $\mathcal{W}^u$, there is an infinity of periodic orbits, and in particular RPOs, such as for instance the one depicted in black in Fig.~\ref{fig:distributions_trajectories}a. As a consequence, the invariant manifolds $\mathcal{W}^s$ and $\mathcal{W}^u$ drive the electron back to the core, and also in the neighborhoods of the RPOs. As a consequence, typical recolliding trajectories resemble one of these RPOs~\cite{Kamor2013}. In addition, $\mathcal{W}^s$ is timing the return of the electrons, as observed in Fig.~\ref{fig:invariant_manifolds}b. The phase space is foliated by $\mathcal{W}^s$. We distinguish different regions with different return times separated by $\mathcal{W}^s$. The return time of the electron depends on which region of phase space it populates at the end of the ramp-up.
\par
For $I > \tilde{I}_{\max}$ where $\tilde{I}_{\max}$ is given by Eq.~\eqref{eq:exact_Imax_definition}, the inner region is depleted during the plateau (see Sec.~\ref{sec:maximum_intensity}), and as a consequence, there are no bounded electrons when the recolliding electron returns during the plateau, and therefore no NSDI are triggered by recollisions. In order to trigger recollisions for $I > \tilde{I}_{\max}$, the return of $\mathrm{e}_1^-$ has to be before the end of the ramp-up, which is what is observed in Fig.~\ref{fig:ionization_time}d--f. These early returns explain the presence of NSDI at large values of the intensity in Fig.~\ref{fig:probability_curves}, when the DI is already saturated. The range of high intensities where these recollisions are possible corresponds to the intensity range where the plateaus of NSDI probability are observed in Fig.~\ref{fig:probability_curves}.
\par
Over all recolliding trajectories leading to NSDI, rare are the ones which undergo more than one recollision, for the set of parameter investigated here. For instance, we consider all recollisions leading to NSDI in CP and LP pulses for intensities lower than $6\times 10^{13}\; \mathrm{W}\cdot \mathrm{cm}^{-2}$. There are about 20 times more recollisions in LP than in CP, but 95\% of the NSDI are with a single recollision in CP, whether this number is about 60\% in LP. We conclude that recollisions in CP are much more effective than in LP, even though they are more rare. 

\subsection{Critical ionization potential and atoms triggering NSDI \label{sec:critical_ionization_potential}}

\begin{figure}
	\centering
	\includegraphics[width=0.5\textwidth]{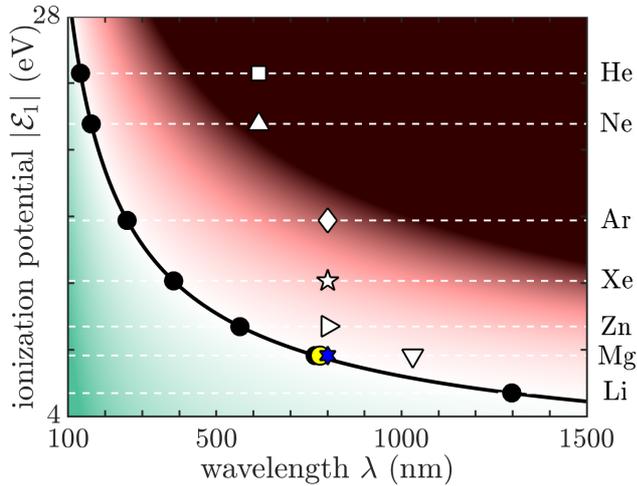}
	\caption{Conditions on the laser wavelength $\lambda$ and the ionization potential $\vert {\cal E}_1\vert$ for which the atom can undergo NSDI in CP (green region) or not (red region). The black line is $\mbox{IP}_c \approx \sigma \omega^{2/3}$ [Eq.~\eqref{eq:NSDI_condition}]. The square ($\square$) is $\mathrm{He}$ and up triangle ($\bigtriangleup$) is $\mathrm{Ne}$ for $\lambda = 614 \; \mathrm{nm}$ reproduced from Ref.~\citep{Fittinghoff1994}. The diamond ($\diamond$) is $\mathrm{Ar}$ and star ($\star$) is $\mathrm{Xe}$ for $\lambda = 800 \; \mathrm{nm}$ reproduced from Ref.~\citep{Guo1998}. The yellow circle ($\circ$) is $\mathrm{Mg}$ for $\lambda = 780 \; \mathrm{nm}$ reproduced from Ref.~\citep{Gillen2001}. The down triangle ($\bigtriangledown$) is $\mathrm{Mg}$ for $\lambda = 1030 \; \mathrm{nm}$, the right triangle ($\triangleright$) is $\mathrm{Zn}$ and the blue hexagon ($\ast$) is $\mathrm{Mg}$ for $\lambda = 800 \; \mathrm{nm}$ reproduced from Ref.~\citep{Hang2020}.}
	\label{fig:NSDI_phase_diagram}
\end{figure}

After ionization, the ionized electron oscillates around the ionic core as a result of the competition between the electric field and the Coulomb potential. These oscillations are well captured by the guiding-center (GC) model~\citep{Dubois2018, Dubois2018_PRE} which is a dynamical reduction of Hamiltonians~\eqref{eq:Hamiltonian_RF_1ae}, valid on long timescales and far from the ionic core. This framework is ideally suited for recolliding trajectories in CP pulses, such as the trajectory depicted in Figs.~\ref{fig:distributions_trajectories}. In Ref.~\citep{Dubois2018_PRE}, we introduced an energy of the GC which conditions the return of the ionized electron $\mathrm{e}_1^-$. If the GC energy is negative, $\mathrm{e}_1^-$ can come back to the core. If it is positive, it ionizes without recolliding. In the adiabatic approximation in the RF, we consider an electron ionizing at time $t_{{\rm i},k}$ through the saddle point of the RF-effective potential. The coordinates of the GC in the RF at ionization time are given by 
\begin{subequations}
\label{eq:guiding_center_coordinates}
\begin{eqnarray}
\tilde{\mathbf{r}}_{g,k} &=& \left[ -\tilde{r}_{{\rm s},k} (t_{{\rm i},k}) - \dfrac{\tilde{E}_k^{\star}}{\omega^2} \right] \hat{\tilde{\mathbf{x}}} ,  \\
\tilde{\mathbf{p}}_{g,k} &=& \omega \left[ -\tilde{r}_{{\rm s},k} (t_{{\rm i},k}) - \dfrac{\tilde{E}_k^{\star}}{\omega^2} \right] \hat{\tilde{\mathbf{y}}} .
\end{eqnarray}
\end{subequations}
The GC energy is given by $G_k = | \tilde{\mathbf{p}}_{g,k}|^2/2 + V_k( \tilde{\mathbf{r}}_{g,k})$, and we consider $V_k ( \tilde{\mathbf{r}} ) \approx - Z_k / | \tilde{\mathbf{r}} |$ by assuming that the guiding center is sufficiently far away from the core. We substitute Eq.~\eqref{eq:effective_amplitude_general_potential_Estar} in Eqs.~\eqref{eq:guiding_center_coordinates}. Using Eq.~\eqref{eq:position_saddle_point_Ip}, the GC energy of the electron at the saddle point of the RF-effective potential at ionization is
\begin{equation}
\label{eq:guiding_center_energy}
G_k = \left( Z_k \omega \right)^{2/3} \dfrac{1 - 2\rho(\mu_k)^6}{2 \rho(\mu_k)^4} .
\end{equation}
At time $t_{{\rm i},k}$ [see Eq.~\eqref{eq:ti_prediction}], $\mathrm{e}_k^-$ ionizes over the barrier. In order to return to the core, the motion of the GC must be bounded, i.e., $G_k < 0$. This condition strongly depends on the time at which the electron ionizes. If the electron ionizes too late, the effective amplitude of the laser is too large and the electron drifts away from the core due to the large amplitude of the potential vector~\cite{Corkum1993}. We denote $\mbox{IP}_c$ the critical ionization potential of the electron such that $G_k = 0$ when $\mathcal{E}_k = - \mbox{IP}_c$. Using Eq.~\eqref{eq:guiding_center_energy} and the condition $G_k = 0$, the critical ionization potential is 
\begin{equation}
\label{eq:NSDI_condition}
\mbox{IP}_c \approx \sigma ( Z_k \omega )^{2/3} ,
\end{equation}
where $\sigma = 3( \mu^{\star}/2)^{1/3}$ and $\mu^{\star}$ is solution of the equation $\rho (\mu^{\star}) = 2^{-1/6}$. Numerically, we have $\sigma\approx 1.85$. It should be noted that a similar calculation in the LF gives $\sigma = 2$. 
If the first ionization potential of the atom, $|\mathcal{E}_1|$, is smaller than $\mbox{IP}_c(\omega)$ for $k=1$, then the atom can experience recollisions, since $G_1<0$. If it is larger, recollisions do not occur, since $G_1>0$. 
Figure~\ref{fig:NSDI_phase_diagram} shows the phase diagram for atoms for which the outermost electron can undergo recollisions as a function of $|\mathcal{E}_1|$ and the laser wavelength. Given the level of approximations, this diagram must be seen as a qualitative indicator for recollisions in CP. Therefore, in principle, recollisions in CP pulses are possible for all atoms, depending on the laser wavelength~\citep{Fu2012,Chen2017}. There exist wavelengths for which certain atoms do not exhibit a substantial probability of recollisions. This regime is of particular interest for Attoclock setup~\citep{Eckle2008}.
\par
\begin{figure}
	\centering
	\includegraphics[width=0.5\textwidth]{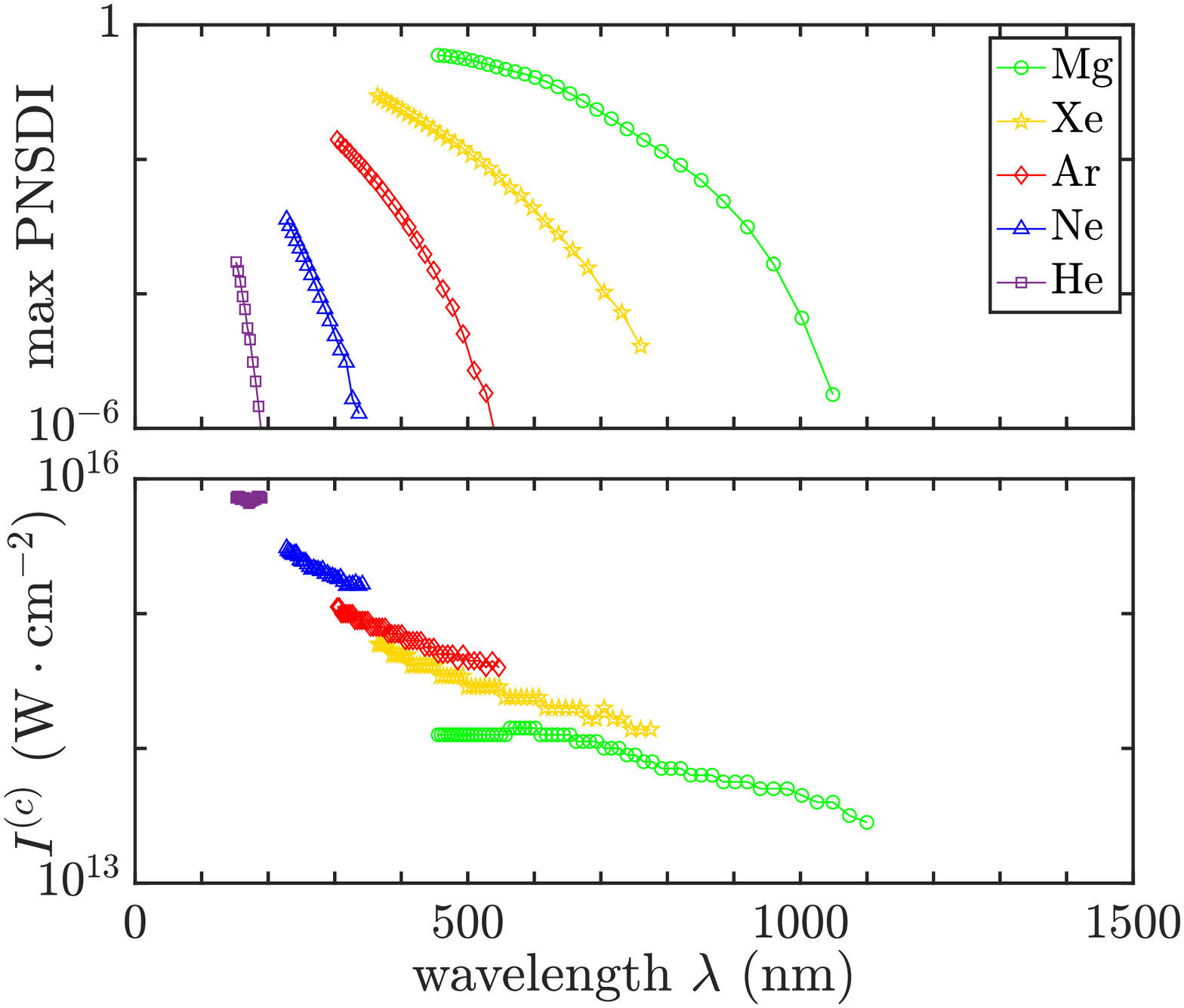}
	\caption{Maximum of the probability of nonsequential double ionization (PNSDI) (upper panel) and intensity $I^{(c)}$ at which the PNSDI is maximum (lower panel) as a function of the laser wavelength $\lambda$, computed from Hamiltonian~\eqref{eq:Hamiltonian_NSDI_LF}. The laser envelope is trapezoidal 2--4--2. The parameters are ($\mathrm{Mg}$) $\mathcal{E}_g = -0.83$ and $a=3$, ($\mathrm{Xe}$) $\mathcal{E}_g = -1.20$ and $a=2$, ($\mathrm{Ar}$) $\mathcal{E}_g = -1.60$ and $a=1.5$, ($\mathrm{Ne}$) $\mathcal{E}_g = -2.30$ and $a=1$, ($\mathrm{He}$) $\mathcal{E}_g = - 2.90$ and $a=0.8$. The maximum of PNSDI becomes smaller than $10^{-5}$ for laser wavelength ($\mathrm{Mg}$) $\lambda_c \approx 1050 \; \mathrm{nm}$, ($\mathrm{Xe}$, $a=2$) $\lambda_c \approx 770 \; \mathrm{nm}$, ($\mathrm{Ar}$) $\lambda_c \approx 510 \; \mathrm{nm}$, ($\mathrm{Ne}$) $\lambda_c \approx 320 \; \mathrm{nm}$,  ($\mathrm{He}$) $\lambda_c \approx 180 \; \mathrm{nm}$.}
	\label{fig:critical_frequency}
\end{figure}

For laser wavelength $780 \; \mathrm{nm}$, $\mbox{IP}_c \approx 7.6 \; \mathrm{eV}$ from the prediction of Eq.~\eqref{eq:NSDI_condition}, i.e., NSDI is expected for atoms whose first ionization potential is around or smaller than $7.6 \; \mathrm{eV}$. This is in agreement with experimental measurements~\citep{Gillen2001, Hang2020} of the DI probability curves of $\mathrm{Mg}$ ($\mathcal{E}_1 = - 7.65 \; \mathrm{eV}$) for which a clear knee structure is observed. In Fig.~\ref{fig:critical_frequency}, we show the maximum of the NSDI probability (PNSDI) as a function of the laser wavelength for $\mathrm{Mg}$, $\mathrm{Ar}$, $\mathrm{Ne}$ and $\mathrm{He}$, computed from Hamiltonian~\eqref{eq:Hamiltonian_NSDI_LF}. The maximum of the PNSDI for a given laser wavelength corresponds to the peak of the dash-dotted curve in Fig.~\ref{fig:probability_curves}. From the prediction of Eq.~\eqref{eq:NSDI_condition}, the critical laser wavelengths are ($\mathrm{Mg}$) $\lambda_c \approx 770 \; \mathrm{nm}$, ($\mathrm{Xe}$) $\lambda_c \approx 380 \; \mathrm{nm}$, ($\mathrm{Ar}$) $\lambda_c \approx 260 \; \mathrm{nm}$, ($\mathrm{Ne}$) $\lambda_c \approx 160 \; \mathrm{nm}$, ($\mathrm{He}$) $\lambda_c \approx 130 \; \mathrm{nm}$. The quantitative discrepancy between the numerical results and Eq.~\eqref{eq:NSDI_condition} are due to the approximations for the derivation of Eq.~\eqref{eq:NSDI_condition}, and to our purely classical calculations, for which the condition $\tilde{H}_k ( \tilde{\mathbf{r}},\tilde{\mathbf{p}},t) \approx \mathcal{E}_k$ before ionization is not necessarily fulfilled. Qualitatively, Eq.~\eqref{eq:NSDI_condition} and the purely classical calculations show that for increasing first ionization potentials, the NSDI probability decreases, as observed in Ref.~\citep{Chen2017}.

The intensity $I^{(c)}$ at which the maximum of NSDI occurs as a function of the laser wavelength is depicted in the lower panel of Fig.~\ref{fig:critical_frequency}. This intensity is such that $I^{(c)} \in [I_{\min} , I_{\max}]$, where $I_{\min}$ and $I_{\max}$ are given by Eqs.~\eqref{eq:minimum_intensity_definition} and~\eqref{eq:exact_Imax_definition}, respectively. It can be roughly and easily estimated by its upper bound, the approximate intensity at which the DI probability saturates $I_{\max}$ given by Eq.~\eqref{eq:bounded_region_Imax_approximation}. It should be noted that the expected values of the intensity $I^{(c)}$ for NSDI depends on the pulse envelope, and in particular on the ramp-up of the laser pulse, as it can be seen on Fig.~\ref{fig:probability_curves}. The intensity at which NSDI can be found increases when decreasing the wavelength, so NSDI might be in the same range as SDI, and hence a knee might not be always visible in a DI probability curve, even though recollisions occur.

\section{Conclusions}

We have investigated some features of the double ionization of atoms subjected to intense circularly polarized laser pulses.   
We have shown that the parameter $\mu \propto |\mathcal{E}_1|^3 \lambda^2$ monitors the amount of nonadiabatic effects for over-the-barrier ionization. For $\mu \gg 1$, the adiabatic approximation, ${\bf E}(t)\approx {\rm const}$, provides accurate estimates, e.g., for ionization times.  
For $\mu \leq 1$, nonadiabatic effects appear. The main nonadiabatic effects identified here are the lowering of the threshold intensity at which over-the-barrier ionization happens and the lowering of the ionization time of the electrons in this regime.
\par
Also, we have studied NSDI processes in CP pulses and determined the crucial role of the laser envelope. We have found that envelope-driven-recollisions~\cite{Dubois2020} are the only recollisions able to trigger NSDI in CP pulses: These recollisions are characterized by pre-ionizations which occur mostly over the barrier and at approximately the same time (depending on the intensity and the laser envelope), early during the ramp-up of the pulse. The pre-ionized electron gains energy from the CP laser pulse due to the variations of the pulse envelope during the excursion in the continuum. It returns to the core at a finite number of return times, depending on the pulse duration, approximately equally spaced by the laser period. This recollision scenario departs significantly from the recollisions in LP fields, where ionizations leading to recollisions occur around each peak of the laser field, i.e., each half a laser period.  
\par
As a result of the nonadiabatic effects, coming from the variations of the laser field on subcycle timescales, and the presence of a laser envelope, all atoms and, in general, all targets are capable of undergoing NSDI in CP pulses, provided that the laser wavelength is small enough (depending on the ionization potential of the target). The conditions to get NSDI with CP pulses are much more restrictive than for LP pulses, so their probability is lower. However the envelope-driven recollisions in CP are more effective than standard recollisions in CP, in the sense that for a larger majority of double ionizations, a single recollision is enough, as compared with LP. 

\section*{Acknowledgments}
We thank Simon Berman, Fran{\c c}ois Mauger and Adrian Pfeiffer for helpful discussions and for sharing their experimental results with us. The project leading to this research has received funding from the European Union's Horizon 2020 research and innovation program under the Marie Sk\l{}odowska-Curie grant agreement No. 734557. T.U. acknowledges funding from the NSF (Grant No.~PHY1602823).

\bibliographystyle{apsrev4-1}
%\bibliography{biblio}
%

\end{document}